\documentclass[
prd,aps,floats,showpacs]{revtex4}
\usepackage{graphicx}
\usepackage{epsfig}

\newcommand{\beq}{\begin{equation}}
\newcommand{\eeq}{\end{equation}}
\newcommand{\be}{\begin{eqnarray}}
\newcommand{\ee}{\end{eqnarray}}

\def\+{\dagger}

\def\simlt{\stackrel{<}{{}_\sim}}
\def\simgt{\stackrel{>}{{}_\sim}}
\begin{document}
\title{Baryon Asymmetry, Dark Matter and \\ Quantum Chromodynamics}

\author{D.H. Oaknin and A. Zhitnitsky}

\affiliation{ Department of Physics and Astronomy, University of British 
Columbia, Vancouver, BC, V6T 1Z1, CANADA}

\begin{abstract}
We propose a novel scenario to explain the observed cosmological asymmetry 
between matter and antimatter, based on nonperturbative QCD physics. 
This scenario relies on a mechanism of separation of quarks and antiquarks
in two coexisting phases at the end of the cosmological 
QCD phase transition: ordinary hadrons (and antihadrons), along with 
massive lumps (and antilumps) of novel color superconducting phase. The
latter would serve  as the cosmological cold dark matter. In
certain conditions the separation of charge is C and CP asymmetric and can
leave a net excess of hadrons over antihadrons in the conventional phase,
even if the visible universe is globally baryon symmetric $B = 0$. In 
this case an equal, but negative, overall baryon charge must be hidden in 
the lumps of novel phase. Due to the small volume occupied by these dense 
lumps/antilumps of color superconducting phase and the specific features of 
their interaction with ``normal" matter in hadronic phase, 
this scenario does not contradict the current phenomenological constrains on 
presence of antimatter in the visible universe. Moreover, in this 
scenario the observed cosmological ratio $\Omega_{DM}\sim\Omega_{B}$ 
within an order of magnitude finds a natural 
explanation, as both contributions to $\Omega$ originated from the same 
physics during the QCD phase transition. The baryon to entropy ratio 
$n_{B}/n_{\gamma}\sim 10^{-10}$ would also be a natural outcome, fixed by 
the temperature $T_f \simlt T_{QCD}$ at which the separation of phases is 
completed.
\end{abstract}
\maketitle

\section{Introduction}

The origin of the cosmological asymmetry between baryons and antibaryons, 
and, more specifically, the origin of the observed baryon to entropy ratio 
$n_B / s \sim 10^{-10}$ ($n_B$ being the net baryon number density in 
hadrons, and $s$ the entropy density) remains a mystery and one of the 
main challenges for particle-cosmology. In order to explain this number 
from symmetric initial conditions in the very early Universe, it is 
generally assumed that three criteria, first laid down by Sakharov 
\cite{Sakharov}, must be satisfied at the instant when the asymmetry was 
generated:
\begin{itemize}
\item{} C and CP are not exact symmetries. 
\item{} Baryon number violating processes exist.
\item{} The processes take place out of thermal equilibrium.
\end{itemize}
In general, nevertheless, all three of these tight conditions can be  
loosened in certain scenarios. For example, C and/or CP asymmetries can be 
generated earlier and not necessarily at the same instant when the 
cosmological baryon asymmetry developed. Explicit baryon number violation 
is not strictly necessary; it has been shown that in certain contexts 
spontaneous violation is able to produce a spatial separation of baryonic 
charge which leaves a positive net number in our local patch of the 
universe. Neither out-of-equilibrium dynamics is a must: in some 
suggested scenarios baryogenesis develops due to spontaneous CPT 
breaking.
 
Here we argue that baryogenesis may be realized at the instant just after 
the cosmological QCD phase transition without explicit violation of baryon 
number, if at some stage in the earlier history large C and CP asymmetries 
have developed homogeneously over the whole visible universe.
Motivated by the cosmological ratio between baryonic and dark matter 
densities, $\Omega_{DM} \sim \Omega_B$ within an order of magnitude, 
which we interpret as an indication of a close relationship between the 
dark matter problem and the origin of the cosmological baryonic asymmetry 
\cite{kaplan}, our proposal features an scenario of baryon charge 
separation in a universe with zero overall baryon charge $B = 0$.
This scenario relies on the idea that baryon/antibaryon 
charge can be stored in dense lumps/antilumps of ordinary light
quarks/antiquarks condensed in color superconducting (CS) phase, 
which would serve as cosmological cold dark matter. In different words, 
we suggest that the observed cosmological excess of hadrons over 
antihadrons may not necessarily be expressed as a net baryon charge $B 
> 0$ in the universe if an equal, but negative, overall charge is 
accumulated in the lumps/antilumps of dense color 
superconducting phase. In this sense, our proposal resembles an old 
scenario discussed in \cite{Widrow} of a baryon "asymmetric" universe 
with zero overall baryon charge $B = 0$, in which the net excess of 
hadrons over antihadrons is compensated by a negative overall baryonic 
charge in an hypothetical new fundamental scalar field hidden in the dark 
sector. Our proposal does not rely on any new fundamental particle, but 
on a new phase of ordinary QCD matter.

This proposal follows the recent observation \cite{Madsen:2001fu, 
Lugones:2003un, qcdball} that chunks of condensed ordinary quarks with 
large baryon charge accumulated in color superconducting phase may become 
at temperatures $T \simlt 0.57 \Delta$, where $\Delta \sim 100$~MeV, 
absolutely stable objects against decay into ordinary nuclear matter. 
Color superconducting (CS) phase is a theoretically motivated novel phase 
of QCD matter that is realized when light quarks are squeezed to a 
density which is a few times the nuclear density and organize a single 
coherent state that condense in diquark channels, analogous to Cooper 
pairs of electrons in BCS theory of ordinary superconductors, see 
original papers \cite {cs_n} and recent reviews \cite{cs_r} on the 
subject. It has been known that CS phase may be realized in Nature in 
neutron stars interiors and in the violent events associated with 
collapse of massive stars or collisions of neutron stars, so it is 
important for astrophysics. We argue here that if such conditions occur 
in the early universe during the QCD phase transition, CS phase may be 
important for cosmology as well.

The possibility that massive lumps of ordinary quarks in a novel state
of QCD matter could form at the end of the QCD phase transition along with 
ordinary hadrons and antihadrons in a cosmological scenario of separation 
of phases was first considered in \cite{Witten}. The proposal noticed that 
the baryon charge hidden in these lumps would not participate in 
nucleosynthesis and, therefore, according to the usual definition 
would not contribute to $\Omega_{B}h^2\simeq 0.02$ in spite of their QCD 
origin. These lumps would serve as ``nonbaryonic" dark matter. 
This original proposal followed the theoretical observation 
\cite{freedman} that at quark densities $\mu \simgt 300$ MeV, larger than 
the strange quark mass and a few times larger than the typical nuclear 
density, addition of strangeness should lower the quark chemical 
potential and ordinary baryonic matter could prefer to settle into a 
novel coherent ground state of three flavours of delocalized quarks,  
known as `strange quark matter', rather than in ordinary hadrons. 

The theoretical discovery of the color superconducting phases of QCD
matter \cite{cs_r, cs_n} has stirred renewed interest on the cosmological 
scenario of separation of phases at the QCD phase transition. The interest 
now focus on the formation of lumps of CS phase  
\cite{Lugones:2003un,qcdball}. It is important to notice that, in the strict 
sense, lumps of color superconducting matter are not the hypothetical 
`strangelets' predicted in \cite{Witten}, in spite of the obvious similarity. 
In color superconducting phases Cooper pairing at high density, rather than 
the addition of a third quark flavour, is the cause responsible for lowering 
the energy and favouring the formation of the novel coherent state: color 
superconducting phases could be made of three flavours of light quarks 
(like CFL phase) or only two of them (like 2SC phases). In addition, 
diquark Cooper pairing in color superconducting phases manifest in a large 
gap $\Delta \sim 100$~MeV in the spectrum of single particle excitations 
\cite{cs_n} of quark matter, which could be a crucial feature to understand 
the phenomenology of these balls/antiballs of condensed matter.
A gapped spectrum of fermionic excitations, on the other
hand, is not a necessary feature of 'strange quark matter'.

We go further with this original idea introduced in \cite{Witten} by suggesting that such dense 
objects could also be configurations with large negative baryon charge 
(antilumps), made of condensed antiquarks in the color superconducting 
phase. As we discuss in next sections in details, in such a form the 
anti-baryon charge could coexist with a net number of hadronic baryons 
without annihilating due to the small volume occupied by
the dense lumps and due to the specific properties of the interaction of 
normal hadronic matter with the lumps in superconducting state. Usual baryons 
(hadrons) could be reflected rather than annihilated when they hit an 
anti-ball with the small energy typical for the present cold universe (when 
$v/c \sim 10^{-3}$). A similar feature is known in the interaction of 
conduction electrons in metals at the interface with conventional 
superconductors \cite{Blonder}. More than that, the anti-baryon charge of 
the anti-ball would not change either the nucleosynthesis calculations because, locked 
in CS phase, it is not available to form nucleons, similar to the 
positive baryon charge locked in CS phase. 

A complete understanding of the mechanisms that produce, in this scenario 
of separation of phases (and charges) at the QCD phase transition, a net 
excess of hadrons over antihadrons in the conventional phase should 
include an explanation of:

1) the mechanisms that lead to the formation at the end
of the cosmological QCD phase transition of lumps and antilumps of
condensed QCD matter/antimatter in color superconducting phase,
coexisting with ordinary hadrons and antihadrons;

2) the mechanisms that avoid the evaporation and dissapearance of the
lumps and antilumps once the temperature of the universe drops well below
the temperature of the phase transition and allow their survival in 
the later colder universe;

3) the mechanisms responsible for the C and CP asymmetric separation of 
the baryonic charge in the two coexisting phases that leaves a net baryon 
number in the ordinary hadronic phase, and how this net excess of hadrons 
is preserved at later times after the separation has been completed and 
antihadrons annihilated;

4) the phenomenological viability of the proposed scenario when confronted
with currently available observational data, in particular, with 
constrains on large amounts of antimatter in the visible universe.

The research on the cosmological scenario of separation of phases 
has focused, since the birth of the idea, on the problems of formation 
and survival of lumps of quark matter at $T_{QCD}$ in a universe that has 
already developed a net baryon asymmetry in the standard sense, as it was 
outlined in the original proposal \cite{Witten}. In spite of all this 
attention, important questions about this scenario still remain open 
\cite{Witten, tres, Sumiyoshi, Lugones:2003un}. The innovative 
aspect of our proposal, in which antilumps of condensed antiquarks form 
along with lumps of condensed quarks and the cosmological baryon asymmetry 
only develops as a result of the net separation of charges, requires a 
re-evaluation of problems 1) and 2) plus an understanding of problem 3), 
which is obviously specific of our proposal. In this paper we shall 
focus, nevertheless, on a phenomenological study of the viability of our 
scenario when confronted with available observational data, problem 4), 
for reasons that are transparent: first, we must be 
convinced that the scenario that we propose is not ruled out; 
second, a complete study of the first three aspects 
is a more complicated task that can be addressed separately. 

At this stage we cannot prove, let us state it clearly, that such dense 
objects of condensed quarks/antiquarks in color superconducting phase do 
form at the end of the cosmological phase transition. Neither, of 
course, we can precisely calculate some important phenomenological 
features of these droplets of condensed phase, such as temperature $T_f 
\simlt T_{QCD}$ at which the formation is completed, mass spectrum of 
resulting balls/antiballs or their number densities. Such 
analysis would require a much deeper understanding of the 
non-equilibrium dynamics of non-perturbative QCD during the cosmological 
phase transition, which is not available at the present time. We 
understand, nevertheless, that we must present some fair evidence 
that issues 1), 2) and 3) can be successfully solved in order 
to make the proposal contained in this paper truly attractive. Indeed, we 
will argue below that a baryon asymmetric cosmological 
separation of phases can be a plausible, and even probable, output of the 
cosmological QCD phase transition.

Our analysis of the phenomenological viability of the scenario, problem 4), focuses on 
the interaction of massive chunks of condensed quarks/antiquarks with 
normal hadronic matter to check that the currently available 
observational data neither rule out this picture nor even impose tight 
constraints on it. We show that large amounts of ordinary quarks and 
antiquarks can exist in the present universe stored in massive 
non-hadronic lumps/antilumps of CS matter, without contravening current 
observational cosmological constraints. Then we argue that in presence of 
large and homogeneous cosmological C and CP asymmetries over the whole 
visible universe at the onset of the QCD phase transition, the 
observational values of both cosmological parameters 
$\Omega_{DM}/\Omega_{B}\simgt 1$ and $n_B/n_{\gamma}\sim 
10^{-10}$ fit very naturally in this scenario.

Let us add before closing this Introducion that many of the essential 
ideas of the scenario advocated here are elaborated versions of some old 
proposals. We have already mentioned that the idea that lumps of ordinary 
quark matter in a novel state, such as strange quark ``nuggets", may play 
the role of dark matter was suggested long ago \cite{Witten}, see also 
original papers \cite{Jaffe} and relatively recent review \cite{Madsen} 
on the subject. It is also an old idea that dark matter may store large 
baryon (or even antibaryon) charge \cite{baryogenesis}. 
The idea that soliton (anti-soliton) -like configurations 
may serve as dark matter, is also not a new idea \cite{Lee}. Maybe, the 
most noticeable example are Q-balls \cite{qball}. As well, the idea that 
baryon density could be inhomogeneous in space while the global baryonic 
charge is zero is old, see reviews \cite{Dolgov1} and references therein. 
The new element of our proposal is an observation that one can 
accommodate all the nice properties discussed previously 
\cite{Jaffe}-\cite{Dolgov1} but  without invoking any new fields and 
particles (except, maybe, from a solution to the strong CP problem in QCD, 
see original papers and reviews in refs.\cite{PQ}-\cite{axion_r}). 
Rather, the dense balls/antiballs of QCD matter are made of ordinary light 
quarks or antiquarks, which however are not in the ``normal" hadronic 
phase, but in the color superconducting phase. 

It is also an important point to be emphasized that the specific structure 
of the objects of condensed quark (antiquark) matter is not really 
relevant for our discussions. In principle, quarks/antiquarks could 
be stored in stable or long-lived metastable ``nuggets" \cite{Witten}, 
CFL-strangelets \cite{Madsen:2001fu}, QCD-balls \cite{qcdball} or any 
other topological or non-topological solitons which have or have not been 
discussed previously, with the condition that they are stored 
in dense color superconducting state. As we already advanced this feature 
might be crucial to understand the phenomenology of balls and anti-balls
and their interaction with surrounding hadronic matter.
 
Our presentation is organized as follows. In Section 2 we review 
the observational data related to the cosmological baryon asymmetry and 
argue in favour of the cosmological scenario of separation of phases as a 
natural framework to explain the origin of the asymmetry and solve the 
dark matter problem. Then we discuss the essential features of the 
interaction of hadronic matter with balls and anti-balls of condensed 
quarks/antiquarks to argue that well-known constraints on absence of 
significative amounts of anti matter in the visible universe can not be 
literally applied to the case when anti-matter is stored in {\it dark} 
massive droplets of the dense color superconductor phase. Moreover, we 
argue that under this hypothesis the observed relation 
$\Omega_B\sim\Omega_{DM}$ is a natural consequence of the underlying QCD 
physics.  Section 3 is  devoted to the 
calculation of the reflection and transmission coefficients of free 
quarks inciding at the interface of the color superconductor. 
These results are used in Section 2 devoted to the observational 
constraints on antimatter in the present cold Universe at 
temperature well below the QCD phase transition. In Section 4 we 
discuss some generic aspects of the charge separation mechanism during 
the QCD phase transition, after noticing that all three Sakharov's 
criteria could be satisfied (in some looser sense) when chunks of dense 
color superconducting matter/antimatter form during the 
phase transition. Finally,  we estimate the fundamental parameter $\eta 
\equiv (n_B-n_{\bar{B}})/n_{\gamma}$ in this scenario.
Section 5 contains our conclusions, where we speculate on possibilities  
to test the suggested scenario.

\section{Baryogenesis vs Baryon Separation }
\subsection{Observations and Phenomenology}
Baryons in hadronic phase make all the directly observable astronomical objects,
from gas and dust to stars and clusters of galaxies, without any
significant trace of antibaryons over a spatial domain that could be
as large as the present horizon, see reviews and some original papers in
\cite{Dolgov:1991fr}. The origin of this asymmetric distribution of 
baryons and antibaryons remains one of the most fundamental open questions 
in cosmology in spite of the great theoretical and experimental efforts it 
has attracted during the last thirty five years, see {\it e.g} recent 
review \cite{Dine:2003ax}.

The baryon-antibaryon asymmetry can be quantified through the ratio 
$(n_B~-~n_{\bar B})~/~n_{\gamma}$, where $n_B$ and $n_{\bar B}$ are, 
respectively, the number densities of baryons and antibaryons and 
$n_{\gamma}$ the number density of photons in the cosmic background. 
Theoretical models predict, and observations confirm, that baryon number 
$n_B-n_{\bar B}$ is preserved in any comoving volume since the time of 
nucleosynthesis, at $T_{nc} \sim 1$~MeV, and probably even earlier, since the 
instant just after the electroweak phase transition, at $T_{ew} \sim 
100$~GeV. Indeed, the anomalous processes which efficiently violate the 
baryon number in the electroweak symmetric phase are effectively 
suppressed soon after the instant of the phase transition when the system 
is in the  spontaneously broken phase. Since the number of photons in a 
comoving volume is also preserved, the ratio $(n_B-n_{\bar B})/n_{\gamma}$ 
remains approximately fixed in physical volumes while the universe is 
expanding and cooling.

If $(n_B-n_{\bar B}) \ll n_{\gamma}$, as it turns out to be the case, the 
early universe is approximately baryon-antibaryon symmetric until    
the QCD phase transition $T_{QCD} \simeq 160 ~$ MeV. This statement 
remains valid even if some asymmetry was generated at some earlier time. 
This is due to the  strong QCD interactions in the quark-gluon plasma 
where massless quark-antiquark  pairs can easily be produced. 
Therefore, $n_B \sim n_{\bar B} \sim n_{\gamma}$ at $T > T_{QCD}$. 
The universe becomes manifestly baryon asymmetric only at the 
temperatures $T \simlt T_{QCD}$, when strong QCD interactions confine
quarks into heavy hadrons, with masses of the order of $m_N \sim 1$~GeV 
which, subsequently, annihilate each other leaving only a small 
excess of hadronic baryons as the remnants, $n_B - n_{\bar B} \sim n_B$. 

The ratio between the number density of this remnant of hadronic baryons 
and the number density of photons in the cosmic background has been 
recently measured with high accuracy, $\eta \equiv n_B/n_{\gamma} \simeq 
6 \times 10^{-10}$ \cite{Spergel:2003cb}. This parameter   can be directly 
related to the ratio $\eta \sim \left(\frac{m_N}{T_{eq}}\right)^{-1}$ 
between the mass of the nucleon $m_N$ and the temperature of 
matter-radiation equality, $T_{eq} \sim 1$~eV. Its precise observational 
value is in excellent agreement with the present cosmological abundances 
of light elements predicted by the standard nucleosynthesis scenario,
see {\it e.g} \cite{Cyburt:2003fe}. 
 
In generic scenarios of baryogenesis at any time earlier than the QCD
phase transition the net asymmetry $n_B - n_{\bar B} = \eta n_{\gamma} \ll 
n_{\gamma}$ should be {\it fine-tuned} to its observational value, in the
sense that $n_B - n_{\bar B} \ll n_B, n_{\bar B}$. 
In most suggested scenarios the need of tuning the ratio $\eta$ 
(when  a natural scale for $\eta$ is absent)  manifests itself by the fact 
that $\eta$ can be made either too large or too small by changing a few 
parameters that in many cases are only loosely constrained. In general, 
fine tuning is not considered as an atractive solution of a problem from 
a theoretical point of view when there is not a satisfactory explanation 
for the tuning itself. This is a very common feature of most scenarios 
for baryogenesis. As we shall see, such kind of fine tuning might not 
be required in the scenario of baryon separation advocated in the present 
work.

This is not the place to review all possible suggested scenarios for 
baryogenesis. However, we wish to briefly comment on the need of 
tuning of some sort in each of the current trends.
For example, in supersymmetric electroweak scenarios the Higgs and stop 
masses must be carefully chosen to lie within narrow intervals to 
generate the observed value of $\eta$. Otherwise, any asymmetry which 
was developed at the electroweak scale will be washed out at the end of 
the phase transition. Other scenarios, like leptogenesis which relies on 
asymmetries generated at higher scales in channels that cannot be erased 
by electroweak physics, need also be carefully tuned. In Affleck-Dine
scenarios the common problem is that too much asymmetry is generated and
must be subsequently diluted. In summary, although the general conditions 
under which the asymmetry could have developed are well-understood, the 
final word about the specific mechanism and physics involved in it still 
remains to be found. 
\subsection{Baryon asymmetry and QCD}

It is of natural interest to explore the possibility that the 
baryon asymmetry may have been generated not before the electroweak phase 
transition (as most scenarios suggest), but later,  at the instant when 
the universe became manifestly asymmetric, immediately after the 
cosmological QCD phase transition. As we have argued, this might be a 
viable scenario relying on nonperturbative QCD physics that, in 
principle, does not require the introduction of any new physics beyond 
the standard model, except maybe for a solution of the strong CP problem. 
Obviously, in absence of baryon violating interactions at $T_{QCD}$ the 
only way to produce a baryon asymmetry is via charge separation. It is to 
a discussion of such a mechanism in the context of a globally 
baryon-antibaryon symmetric universe, $n_B^{tot} = n_{\bar B}^{tot}$, to 
which we now turn. In such a scenario the ratio $\eta$ is fixed by the 
energy scale of the physics involved and the problem  of fine tuning 
mentioned above is automatically resolved, see below.

This scenario for baryogenesis should be examined in the context 
of recent observations that confirm that baryonic (to be precise: the 
hadronic) matter contributes only a fraction of the matter content of  
the universe, while a much larger fraction is made of some 
unknown form of {\it dark} matter which is not apparent to detection 
through electromagnetic radiation, $\Omega_B \sim \frac{1}{6}\Omega_{DM}$ 
\cite{Spergel:2003cb}. In this context, a net number of hadronic baryons 
$n_B - n_{\bar B}~\neq~0$ can be generated at the QCD phase transition if 
some mechanism exists that separates baryon and antibaryon charges  and 
stores an excess of the latter in compact objects of 
non-hadronic, color superconducting phase discussed  above. 
The non-hadronic objects, which however carry a large baryon 
charge $\pm B$, have heavy mass $M_{DM}$ and  would contribute, instead, 
to the "non-baryonic" cold {\it dark} matter of the universe,
according to the standard definition of the ``non-baryonic" dark matter. 
Once the phase transition has been completed hadrons will annihilate 
anti-hadrons leaving at the end only the excess of the former plus an 
equal negative excess of baryonic charge in the objects of condensed 
phase. In this case a total baryon-antibaryon annihilation after the QCD 
phase transition would be avoided, even though $B = 0$, as we will discuss 
in detail in next 
sections. The separation of phases must be completed before the 
nucleosynthesis, at $T_{nc} \sim 1$~MeV, such  that only the surviving 
hadronic baryon number $n_B = \eta n_{\gamma}$ participates in the 
composition of light nuclei. 

Let us assume that this hypothetical scenario is correct and dark 
matter indeed consists of heavy objects made of quark matter in color 
superconducting phase. What phenomenological consequences can we derive 
from this assumption? 
As the total baryon number is conserved and hadronic baryons have 
charge $+1$, the net number density of non-hadronic QCD-balls 
should be

\begin{equation}
\label{1}
{\widetilde n}_{\bar{B}} - {\widetilde n}_{B} = \frac{1}{B}( n_B- 
n_{\bar{B}}) \simeq \frac{1}{B} n_B,
\end{equation}
where we introduce notation  ${\widetilde n}$ describing the density
of dark matter heavy particles which carry the baryon charge in a hidden 
form of the diquark condensate (CS phase) rather than in form of free 
baryons. 
Let then consider the ratio of $dark ~matter~ number~ density 
\equiv {\widetilde n}_B + {\widetilde n}_{\bar B}$ to $baryon ~number~ 
density \equiv n_B$. By definition,

\begin{equation}
\label{r1}
\left(\frac{dark ~matter~ number~ density}{baryon~ number~ density}\right)
\simeq \frac{m_N\Omega_{DM}}{M_{DM}\Omega_B }.
\end{equation}
The $dark ~matter~ number~ density$ could be naturally estimated, without 
any {\it fine-tuning}, to be 

\begin{equation}
\label{2}
{\widetilde n}_{\bar{B}} + {\widetilde n}_{B} =\# ({\widetilde n}_{\bar 
B} - {\widetilde n}_B),
\end{equation}
where $\#$ is some numerical factor $ \simgt 1$, if the excess
$({\widetilde n}_{\bar B} - {\widetilde n}_B)$ is of the same 
order as the number densities ${\widetilde n}_{\bar B}$ and  ${\widetilde 
n}_B$. In fact, the excess $({\widetilde n}_{\bar B} - {\widetilde n}_B)$ 
is indeed expected to be of order ${\widetilde n}_{\bar B}, ~{\widetilde 
n}_B$ if the universe  is largely C and CP asymmetric at the onset 
of formation of the condensed balls. Then, the l.h.s. of the ratio 
(\ref{r1}), can be estimated to be 

\begin{equation}
\label{3}
({\widetilde n}_{\bar B} + {\widetilde n}_B)/n_B 
= \# ({\widetilde n}_{\bar B} - {\widetilde n}_B)/n_B  \simeq \frac{\#}{B}
\end{equation}
according to eqs. (\ref{1},\ref{2}). 
Consequently, from (\ref{r1},\ref{3}) we obtain 

\begin{equation}
\Omega_{DM}/\Omega_B \simeq (\#/B) \times (M_{DM}/m_N).
\end{equation}
Now, if one demands $\label{input} B \simeq (M_{DM}/m_N)$, which 
is a condition for the  stability of the balls, one can immediately
derive $\Omega_{DM} / \Omega_B \simeq  (\#) \simgt 1$. The point we want to 
make is: our assumption that the dark matter is originated at the QCD 
scale from ordinary quarks fits very nicely with $\Omega_{DM} / \Omega_B 
\simgt 1$ within the order of magnitude, provided that separation of 
baryon charges is also originated at the same QCD scale. Generally, the 
relation $\Omega_B \simlt \Omega_{DM}$, within one order of magnitude, 
between the two different contributions to $\Omega$ is difficult to 
explain in models that invoke a dark matter candidate not related to the 
ordinary quark/baryon degrees of freedom. 

\subsection{Cosmological scenario of separation of phases}

The possibility that massive lumps of ordinary quarks in a novel state
of QCD matter could form at the end of the QCD phase transition along 
with ordinary hadrons and antihadrons and serve as cold dark matter 
in a cosmological scenario of separation of phases was first considered 
in \cite{Witten}. The idea has attracted since then a lot of effort 
\cite{tres, Sumiyoshi, Lugones:2003un,qcdball}, but  
crucial questions about this scenario have remained open. Namely, 
the mechanisms of formation of the lumps and their survival once the 
temperature of the universe drops well below the temperature of the phase 
transition are not yet completely understood. 
While the formation of dense lumps during the phase transition is beyond 
the scope of the present paper, we nevertheless want to present some 
arguments to show that within a broad consideration of the thermodynamical  
parametrical space at the onset of the cosmological QCD phase transition 
a separation of phases that leaves a net baryon number in the 
conventional hadronic phase  is a plausible, and even probable, 
output of the transition.

In the original proposal \cite{Witten} it was discussed that lumps of 
quark matter would form adiabatically in a first order QCD phase 
transition in a two stages process. First, large bubbles of quark matter 
would form at $T_{QCD}$ in thermal and chemical equilibrium with the 
expanding hadronic phase. In a second stage these bubbles of quark matter 
would shrink, while delivering heat and entropy to the surrounding 
medium. The fate of the bubbles of quark matter depends largely on the 
leading mechanism of heat/entropy release. If it mainly proceeds through 
baryon evaporation at the surface of the bubbles, they will lose their 
mass and finally they will dissapear. On the other hand, if the 
heat/entropy release proceeds much faster than baryon number evaporation 
(for example, through emission of Goldstone modes), then a large baryon charge 
could get trapped within the walls of the shrinking bubble in thermal, 
but not chemical, equilibrium with the surrounding hadronic phase. In 
this case, the shrinking walls squezze the trapped baryon 
charge until the mounting Fermi pressure from within stops further 
shrinking. At this point the quark matter density within the bubble could 
reach the density threshold beyond which it organizes the diquark 
condensate. 

A clear drawback of this mechanism is the requirement of a first order 
transition: current theoretical analysis and numerical simulations of the 
QCD phase transition seem to show that the transition is not first order, 
but a crossover, unless the strange quark mass is rather light. 
A different problem of this scenario has to do with the mechanisms that 
could avoid the evaporation of the balls of quark matter, in the case they 
indeed form, once the temperature of the universe drops well below the 
temperature of the QCD phase transition. Balls of quark matter can 
coexist in thermal equilibrium with the surrounding hadronic environment 
at temperatures close to the phase transition. It has been known that 
lumps of quark matter with large baryon charge $B \simlt 10^{40}$ would 
also be thermodynamically stable against decay into ordinary nuclear 
matter at temperatures $T \simlt T_s$, with $T_s = 0.57 \Delta 
\sim 60$~MeV, when they settle in color superconducting phase: the typical 
binding energy is estimated in the literature to be of the order of 
$\sim 50$~MeV, per baryon. But, 
at intermediate temperatures $T_s \simlt T \simlt T_{QCD}$ the lumps of 
quark matter would be unstable and should evaporate before they cool off 
to the temperature $T_s$ at which they become stable configurations, see 
for example \cite{Lugones:2003un} and references therein. To solve these 
problems one can imagine a scenario involving additional fields which make the formation
of such lumps much more likely to occur. One particular example would be
the frustrated collapse of axion domain walls immediately after the completion 
of the QCD phase transition \cite{qcdball}. This mechanism could work even if 
the QCD phase transition is second order or a crossover and the stability of the
droplets of CS matter formed under the squeezing pressure of 
collapsing axion domain walls would supposedly be immediate if they form 
with sufficiently large baryon charge. 

Either in its original form \cite{Witten} or in the elaborated version 
\cite{qcdball}, the mechanism of cosmological separation of phases 
through squezzing of quark matter within the walls of shrinking bubbles 
assumes, in a subtle way, that baryogenesis (in the standard sense of 
generation of net baryon number density $n_B/s \sim 10^{-10}$) has 
already happened at some earlier stage before the QCD phase transition. 
Hence, when the initial bubbles of quark matter form in chemical 
equilibrium with the surrounding hadronic phase at $T_{QCD}$ they already carry a net 
baryon number $n_B/s \sim 10^{-10}$, which is subsequently squezzed to 
higher densities $n_B/s \sim 1$ when the bubbles shrink. In order to 
implement a similar mechanism in our proposal, in which the average 
baryon number in the plasma is exactly zero $B = 0$ and the baryon 
asymmetry of the universe develops only as a consequence of the 
cosmological separation of phases, we should address, in addition, how 
the bubbles of quark matter can trap, 
while shrinking, either a positive or negative large baryon charge. This 
could happen, in particular,  if the walls that contain the collapsing bubbles of quark 
matter show different penetrability for baryons and antibaryons. A 
typical example of such feature is the  CP asymmetric axion domain walls, as suggested 
in \cite{qcdball}, assuming the QCD plasma is already largely C asymmetric due to weak
interactions.

After this brief recall of the current status of ideas on the issues of 
cosmological formation and survival of lumps of quark matter, we wish to 
add a novel observation that could be of significative relevance for the 
understanding of the cosmological scenario of separation of phases at the 
QCD phase transition and the genesis of the cosmological baryon asymmetry 
and the dark matter content of the universe. First, let us notice that 
current lattice simulations that explore the dynamics of QCD matter at 
finite temperature are carried, due to analytical needs, for a plasma 
in which the quark chemical potential is exactly or very close to zero 
$\mu \simeq 0$. This setup is thought to be a good first approximation to 
the real QCD phase transition, because at $T \simgt T_{QCD}$
the universe is closely baryon antibaryon symmetric, $n_B - n_{\bar 
B} \ll n_B, n_{\bar B} \sim n_{\gamma}$, even in the case when 
baryogenesis (in the standard sense) has already occurred. This 
complacence is, notwithstanding, unjustified. In the quark gluon plasma 
at temperatures above the temperature of confinement there coexist three 
flavours of light quarks, $u$, $d$ and $s$, and their corresponding 
charge conjugates. Each fermion can appear with two possible chiral 
orientations and two possible values for their helicity. That means that 
the quark plasma is characterized, in general, by twelve different 
chemical potentials $\mu_i$, as color gauge symmetry guarantees that 
chemical potential for the three color orientations of each sub-specie 
are all equal. If the average baryon number is zero or close to zero, $B 
\simeq 0$, then the twelve different potentials are constrained to add up 
to zero, $\sum_ i \mu_i \simeq 0$. It is necessary, in addition, that 
$\sum q_i \mu_i = 0$, in order to keep electrical neutrality. Otherwise, the 
twelve chemical potentials can be considered as an unknown set of free
parameters that fully characterize the thermodynamics of the cosmological 
QCD plasma. These parameters can, in principle, take values in a very 
broad range, only constrained by Maxwell-Boltzmann equations of thermal 
equilibrium and the mentioned equations on charge conservation. In the 
simplest, but particular, case when all these 
chemical potentials are zero, the cosmological QCD plasma is C and CP 
symmetric. In presence of some non zero chemical potential C invariance 
is broken. CP invariance is also broken, unless the chemical potentials for 
a given sub-specie and its parity conjugate are exactly opposite. 
In summary, even when the net baryon charge of the universe is zero $B = 
0$ chemical potentials for the different quark sub-species in the 
cosmological QCD plasma can be very large. Indeed, large quark chemical 
potentials in the cosmological QCD plasma should be naturally expected, 
even when globally constrained to $\sum_i \mu_i \simeq 0$,
because C conjugation is largely violated by weak interactions. 
Moreover, cosmological CP violation is also generally believed to happen 
before the QCD phase transition, see below. For the 
development of non-zero chemical potentials the violating interaction must
proceed in conditions of non equilibrium. 

The dynamics of the QCD phase transition in presence of 
large quark chemical potentials $\mu_i \neq 0$ might be significatively 
different from the dynamics observed in current lattice simulations of the 
C and CP symmetric scenario when all $\mu_i \simeq 0$.
In fact, a quick glance at the 
phase diagram of QCD at non-zero $\mu_i$   suggests that in presence of 
large chemical potentials a direct phase transition from the quark-gluon 
plasma into lumps and antilumps of color superconducting phase (avoiding 
the transition into conventional hadronic phase), or a mix of hadrons 
and antihadrons coexisting with lumps and antilumps, could be a plausible
outcome of the phase transition at $T_{QCD}$. In this case, the QCD phase 
transition could happen at temperatures as low as $T \simlt 0.57 \Delta 
\sim 60$~MeV, depending on the specific values of the chemical potentials 
of each quark sub-specie, and the stability and survival of the formed 
lumps would be immediately guaranteed. In addition, we will discuss below, 
in subsection III D, that once a large baryon fluctuation in the condensed  
CS phase is formed at temperatures just below the phase transition it could tend 
to grow by trapping more baryons from the surrounding "normal" phase.

Although these arguments, of course, do not prove that cosmological 
separation of phases indeed do happen at the QCD phase transition, they do 
show that a wide range of plausible cosmological scenarios could lead to 
a succesfull separation of phases. Obviously, our proposal is not 
committed, as we have said, to any of these particular mechanisms or any 
others that could be found suitable. 

\subsection{Antimatter as dense  color superconductor}
It is important to remark here that bounds that tightly constraint the 
presence of significant amount of antimatter in regions of the universe 
of different size scales are mainly derived from the phenomenological 
signatures of electromagnetic matter-antimatter annihilation processes 
\cite{Dolgov:1991fr}. These bounds do not strictly apply to the presence 
of antimatter stored in color superconducting phases if this 
kind of objects do not easily annihilate. Or in more precise words, if 
the rate of annihilation is highly suppressed. Here we want to use 
the physical picture of conventional superconductors to qualitatively 
explain why normal hadronic matter may not be annihilated, but 
reflected, by objects made of color superconducting antimatter. 
More detailed calculations which support the qualitative arguments presented in 
this section are presented in next Section 3.

The peculiarities of the scattering process of conducting electrons  
from the metal at a superconductor junction are known to be a 
consequence of the energy gap $\Delta$ in the spectrum of single particle 
excitations of the superconductor above the Fermi surface. The basic 
features of the interaction are discussed in a unified framework using 
Bogoliubov - de Gennes equations, see \cite{Blonder}. The observed 
phenomena can be explained in the following simple way. The conducting 
electrons of the metal, modelled as a gas of free fermions, inciding at 
the interface with energies much smaller than the energy gap $\Delta$ 
cannot go through the interface simply because there is no any 
kinematically available state in the superconductor for a single 
electron. Therefore, conduction electrons from the metal inciding at the 
interface must, in principle, be reflected backward into the 
metal. This process is known in the literature as normal reflection at 
the metal-superconductor interface. There is, nevertheless, a peculiar 
way for an incident low energy electron to propagate into the 
superconductor  known as Andreev reflection \cite{Andreev}: the incident 
electron excites an additional second electron from the metal ``sea"  to 
form a Cooper pair  (which can be excited in the superconductor without 
surpassing any gap barrier) and  leaves a hole that propagates backward 
into the metal. In terms of particle physics concepts, Andreev 
reflection is an example of particle antiparticle pair creation, 
in which the particle (electron) joins the incident electron and together
propagate inside the superconductor as a Cooper pair, while the 
anti-particle (hole) propagates backward. Normal reflection of the 
conduction electron as an electron and Andreev reflection of the electron 
as a hole compete and their relative probabilities are determined by the 
properties of the interface. In presence of a high potential barrier 
$V(x) =  \nu \delta(x)$ at the interface which separates the metal and 
superconducting phases, classic normal reflection happens with 
probability close to one, ${\cal P}(e \rightarrow e) \simeq 1$, while the 
probability for Andreev reflection is effectively zero, ${\cal P}(e 
\rightarrow {\bar e}) \simeq 0$ \cite{Blonder}. This limit is known in the 
literature on ordinary superconductors as classic interface, while the 
opposite situation when there is no barrier and Andreev 
reflection is most probable is known as metallic interface.

Now, let consider an antiparticle (hole) from the metal inciding at 
the interface with the superconductor. The energy gap
$\Delta$ in the spectrum of single particle excitations in the bulk of the 
superconductor kinematically forbids a low energy, $E < \Delta$, incident 
hole to penetrate the bulk of the condensed phase and annihilate with a 
paired electron, because the second electron of the couple should be 
promoted above the energy gap. Annihilation, therefore, cannot happen in 
the bulk of the superconductor. It could, nevertheless, happen at the 
interface if the resulting unpaired electron is expelled from the 
superconductor into the metal. In order to estimate the probability of 
this annihilation process, in which an incident hole is repelled as an 
electron off the superconducting phase, we notice that this process
is nothing but the time reversal of the Andreev reflection discussed in 
the paragraph above. As time reversal symmetry is preserved in common 
superconductors, the probability for particle-antiparticle~(hole) 
annihilation at the metal-superconductor interface must be equal to the 
well-known probability for Andreev reflection (particle-hole pair 
production). As we have remarked before, in presence of a 
high potential barrier $V(x) = \nu \delta(x)$ Andreev reflection is very 
suppressed at low energies $E \simlt \Delta$, which in turn implies that 
also electron-hole annihilation must be highly suppressed. These 
conditions offer us an explicit example of a normal-superconductor 
interface in which particle antiparticle~(hole) annhilation can be avoided. To 
be more precise, the cross section for particle hole annihilation 
when the former is stored in superconductor phase can be largely suppressed 
when compared to the ordinary annihilation cross section in the metal. In 
these conditions the incident hole from the metallic phase is necessarily 
reflected back into the metal as a hole.

We want to use this experience gained from the analysis of conventional 
superconductors for a qualitative understanding of the 
interaction of ``normal" hadronic baryons with chunks of quark matter 
or antimatter in CS phase, characterized by a very large gap $\Delta \sim 
100$~MeV in its spectrum of single particle excitations. We note that the 
interaction at temperatures much below $T_{QCD}$ of hadrons inciding at 
the chunks of superconducting phase with kinetic energies much 
smaller than the gap $E \ll \Delta$ is not asymptotically free and cannot 
be simply described as a quark-CS phase interface in terms of the form 
factors of quarks in the hadron. In spite of that it is convenient and 
easier to discuss first the scattering of constituent quarks 
by balls/antiballs of CS phase. Incident quarks 
should be reflected off the bulk of the color superconducting
phase, either baryonic or antibaryonic, with reflection coefficient close
to one, similar to the case of high potential barrier interface in ref. 
\cite{Blonder}, if they do not carry enough energy to penetrate the 
superconductor. This discussion is presented with explicit 
calculations in next section. In this case, the high potential barrier at the 
interface is produced by the large difference in the vacuum 
energies, or bag constants, that characterize the two phases in contact.

It must be remarked that a proper analysis of annihilation processes of real
incident quarks (rather than holes) at the interface with the antibaryonic CS phase
must include the available mass at rest of the interacting particles, which
could radically change the simple arguments presented here.
In ordinary superconductors slow incident positrons can annihilate in the
bulk and promote the unpaired electron above the gap because the mass at rest
of positrons/electrons $\sim 0.5$~MeV is many orders of magnitude larger
than the energy gap $\Delta \sim 10^{-3}$~eV in the superconductor. 
The situation is quite different for color superconducting quark matter. 
The gap $\Delta \sim 100$~MeV in CS phase is of the same order of magnitude than the typical mass 
of constituent quarks in the hadronic phase. Moreover, the 
effective masses of current quarks in the bulk of the dense CS phase are largely reduced 
by strong interactions. On similar grounds a large suppression 
factor 20-30 in the annihilation rate of antinucleons in dense ordinary nuclear 
matter (with respect to nucleon antinucleon annihilation in vacuum) 
was predicted in \cite{Mishustin}, after noticing that the lower effective 
masses reduce the available phase space for annihilation. 
The suppression factor should be expected to be much larger in the denser 
CS phase where the effective masses are even lower and the available phase space is 
further restricted by the large energy gap $\Delta$. A detailed calculation of these 
effects is out of the scope of this paper. We wish simply to highlight that 
annihilation cross section of incident quarks can be largely
suppressed when the antimatter is stored in dense droplets of the gapped dense color 
superconducting phase. As the individual quarks that form an inciding hadron 
have different quantum numbers than the Cooper pairs in the CS phase, we argue 
that for hadron to penetrate/annihilate the condensed phase the individual 
quarks need to cooperate coherently and, in consequence, the process is 
kinematically further suppressed. 

Therefore, QCD anti-balls and balls, if formed, could behave as very 
stable, massive and solid soliton-like objects that carry large, negative 
or positive, baryon number and reflect off usual hadrons inciding on 
them. This feature, as we advanced in the Introduction, plays 
a crucial role in the explanation of the phenomenology of these 
``non-baryonic" droplets and how they interact with the surrounding  
hadronic matter after the universe cools down to temperatures well below 
$T_{QCD}$. Such peculiar interaction features would imply that if the 
ball or antiball with small velocity $v/c\sim 10^{-3}$ enters a large 
massive object made of usual hadronic matter, for example the Earth, it 
will not necessarily decay by exploding. Rather it will go through the Earth
leaving behind the shock waves. 
Direct searches of non-topological Q-balls, which have phenomenologically 
similar interaction features with ordinary matter, have been reported in
\cite{Arafune}. At this point the search cannot rule out the existence in
the present universe of a density of heavy solitons (or antisolitons) 
with $|B| \simgt 10^{17}$ and $M_B \simeq |B| m_N$ in number 
large enough to explain the content
of dark matter in the universe, ${\widetilde n}_{\bar B},{\widetilde n}_B 
\sim \frac{\eta}{B} n_{\gamma}$. In this context,
it is tempting (but yet futile 
due to the short statistics) to identify the recent  
seismic event with epilinear source\cite{Teplitz}  as the process which 
involves the dark matter particle, such as the balls of CS matter.

The physical arguments presented in this section lead to the 
conclussion that the cross section for annihilation between 
matter and antimatter segregated in different phases may be highly 
supressed in comparison to the standard cross section in vacuum. As we 
have already noticed, this point is extremely important for the 
explanation of the phenomenology of color superconducting QCD anti-balls 
in the presence of surrounding ``normal" baryons in the cold universe. 
Moreover, if the annihilation cross section is strongly suppressed as 
discussed above large amounts of antimatter could exist inside the 
present visible universe, thus avoiding current observational 
constraints.

We wish to add one more comment to the last statement. The rate for 
matter antimatter annihilation is proportional not only to the cross 
section for the interaction, but also to the flux of collisions.
In our previous discussions we have presented arguments which show
that the cross section for matter antimatter annihilation can 
be largeley suppressed when the antimatter is stored in CS, non-hadronic 
phase. Now, we want to add that, in addition, the number of quark 
antiquark encounters per unit time and per unit volume when antimatter 
is stored in dense chunks of CS phase will be also largely reduced with 
respect to the case when both are in the same ordinary hadronic phase.
Therefore, matter antimatter annihilation rate in the hypothetical 
scenario we have introduced is very supressed due to both of 
the two factors: suppression in the annihilation cross section and, moreover, 
suppression in the net flux of interactions. This favours the possibility 
that large amounts of antimatter could be stored in CS phase without 
violating current observational constraints.

Indeed, we can estimate the total number of collisions between ordinary 
hadrons and the dense balls of color superconducting phase in a Hubble time.  
The average number density of balls/antiballs in the halo of the Galaxy 
is, ${\widetilde n}_B\sim \frac{1}{B} \frac{\rho_{DM}} {1 GeV}$ 
with $\rho_{DM} \sim 0.3 GeV/cm^3$ and, thus, the number of collisions per 
hadron per unit time is given by
\beq
\label{insert_1}
\frac{dW}{dt}=4\pi R^2{\widetilde n}_{B}v  \simeq 
4\pi R^2 \frac{1}{B} \frac{\rho_{DM}}{1 GeV}v ,
\eeq
where $v \sim 10^{-3} c$ and $R\simeq 0.7\cdot 10^{-13}B^{1/3}cm$ is the
typical radius of the balls/antiballs.  
Even if  annihilation were 100\% efficient (which is not the case as argued 
above), the resulting lifetime for hadron annihilation is much longer, 
$H \frac{dW}{dt} \sim 0.2 \times B^{-1/3} \ll 1$, than the present Hubble time, $H\simeq 3\cdot 10^{17}s$, 
if one uses already existing observational constraint on such kind of objects, 
$B \geq 10^{18}$ \cite{qcdball,Arafune}. 

The same result can be alternatively presented in terms of the fraction
of the baryon charge of an antiball that would be annihilated in a Hubble time.
The number  density of hadrons in the ordinary phase is, on average, 
$n_B\sim\frac{0.15\rho_{DM}}{1 GeV}$. Thus, the number of collisions per unit time 
in presence of a single QCD ball is given by
\beq
\label{insert_1}
\frac{d{\widetilde W}}{dt}=4\pi R^2n_{B}v  \simeq 
 4\pi R^2\frac{0.15\rho_{DM}}{1 GeV}v.
\eeq
Even if the annihilation is 100\% efficient,  the total (anti) baryon charge from anti QCD ball $\Delta B$
which will be  destroyed by such annihilations during a Hubble time does not exceed  
\beq
\label{insert_2}
\Delta B \simeq \frac{d{\widetilde W}}{dt} \cdot H\leq 0.1 B^{2/3},
\eeq
per QCD ball with charge $B$. This represents an exceedingly small
part of the QCD ball, $\Delta B/B\sim 0.1 B^{-1/3}$ for sufficiently large  $ B$. 
If one uses already existing constraint on such kind of objects,  $B \geq 10^{18}$,  
one concludes that there is no obvious 
contradiction of the suggested scenario with present observations.
 
Obviously, complications derived from the non-uniform
distribution of dark matter over astrophysical scales may modify this rough 
estimation. However, we do not expect that this complications
will make qualitative changes to the arguments presented above. 
We should add, that more refined observations such as measurement of a single 
$511 KeV$ line from the bulge (as observed by INTEGRAL), or galactic diffuse 
spectrum in the $GeV $ range (`` GeV"   excess)  as observed by EGRET 
could be associated with annihilation of the visible matter with QCD 
(anti)balls. The corresponding analysis will be presented
somewhere else \cite{511}. 
 
The possibility that small domains of antimatter of astrophysical or 
cosmological size may exist in the universe and they do not contradict 
current observations was discussed earlier \cite{Khlopov1} and it was even  
suggested that these antimatter domains could evolve into condensed 
antimatter astrophysical objects \cite{Khlopov2}. Our proposal 
is not based on antimatter domains of such a big astrophysical size, 
but on droplets of macroscopic size of CS phase, and we wish to notice 
that when antimatter is stored in the dense phase small volumes can, 
nevertheless, account for a large amount of antimatter.

In summary, we have argued in this section that there might not be
any phenomenological contradiction to the proposed scenario of the 
universe with zero total baryon charge, which is however distributed in 
two phases. The visible content consists of `` normal baryons "
in the hadronic phase with positive baryon charge, while the dark content is in the color 
superconducting phase( which contains both: positive as well as negative baryon charges). 
 
\section{Reflection and Transmission Coefficients}

 We consider in this section the 
scattering of quarks and antiquarks off a surface separating color 
superconducting ~(CS) and 
plasma phases and calculate their reflection and transmission 
({\cal R}\&{\cal T}) coefficients. This calculation involves 
only perturbative dynamics of quarks at the Fermi surface, $\mu \sim 400$~MeV,
while all nonperturbative physics is assumed to be parameterized
 by the diquark condensate.
In order to avoid complications that may not be 
relevant at this stage the scattering problem is reduced to a one 
dimensional calculation of {\cal R}\&{\cal T} coefficients,
assuming that the typical size of the ball of condensed matter is much 
larger than all other scales involved in the scattering process.

Our goal is twofold. First, we want to demonstrate that under certain 
conditions the reflection coefficient is exactly one (complete 
reflection) if the energy of the incoming quark or antiquark is smaller 
than the energy gap $\Delta$ in the spectrum of single particle 
excitations of the CS phase. This is the main goal of this section. 
This feature, as we have remarked in previous sections, is very important 
for understanding the phenomenology of balls and anti-balls as dark 
matter during the epoch when the universe has cooled down to temperature 
well below $\Delta \sim T_{QCD}$. Such a phenomenon of total reflection 
of fermions off a superconducting region is well known in condensed 
matter physics in the interaction of conduction electrons of a metal at a 
high barrier junction with a conventional superconductor \cite{Blonder}. 

Our second goal is to demonstrate that at larger energies
(such energies are typical when the temperature is high, $T\simeq 
\Delta$) there can be a net transport of baryon number through the
interface into the CS phase. This feature is important to understand 
the mechanism of charge separation that could lock quarks/antiquarks in 
the form of balls/anti-balls during the phase transition. We wish to 
remark again that we do not pretend to prove with our simple discussion 
that separation in CS objects does in fact occur during the cosmological 
phase transition. We only pretend to show that the conditions needed for 
the separation to happen can be fulfilled: this discussion must be 
understood only as an indication that separation can happen. 
	
We should remark here that a similar scattering problem of particles 
off the interface region separating CS and hadronic phases was   
discussed previously in \cite{Sadzikowski:2002ib}. The analysis of these 
papers was motivated by the physics of neutron stars where CS phase is 
very likely to develop. Our context here is very different: we study the 
interactions of heavy objects made of condensed CS matter with the {\it 
gas} of hadrons that surround them during and after the cosmological QCD 
phase transition. However, the technique developed in  
\cite{Sadzikowski:2002ib} turned out to be very usefull and will be widely 
used in the analysis which follows. The equations that describe the
interface are basically the same Bogoliubov - de Gennes equations used in 
the classic reference \cite{Blonder} to describe the ordinary metal 
superconductor junction, with the remark that reference 
\cite{Sadzikowski:2002ib} works with relativistic Dirac equation and 
\cite{Blonder} works in the non-relativistic Schrodinger framework. The 
specific features of the setup 
in  \cite{Sadzikowski:2002ib} for describing the interface of phases in 
neutron stars  match those we need for studying the process of formation 
of chunks of condensed color superconducting matter inside dense clouds of 
quark plasma during the QCD phase transition when $T \sim \Delta$.
On the other hand, the features of the setup that appropriately 
describes the interface of CS and hadronic matter in the cold universe at 
temperatures well below the phase transition $T \ll \Delta$ are 
significatively different, because in such environment the density of 
quarks outside the ``nuggets" of CS matter is much lower than the 
density of baryon charge inside of the ``nuggets"
\footnote{ Of course, there is no any physical jumps
in chemical potentials between these two phases.
 However, we model the interface region (which
probably includes a mixture of hadronic matter as well as nuclear matter) as
a sharp $\theta(x)$ function assuming  that the inverse width of
the interface region is much larger than any other scales of the problem.}. 

Let us now start our detailed discussion with the effective lagrangian 
\begin{equation}
\label{lagrangian}
{\cal L}={\bar \psi}^a_i \left(i \partial_{\mu}\gamma^{\mu} - m 
+ \mu\gamma^0 \right) \psi^a_i + 
\{\Delta^{ab}_{ij}\left(\psi^{a 
T}_i C\gamma^5 \psi^b_j\right)^{\dagger} + h.c.\},
\end{equation}
which describes the relevant fermionic degrees of freedom at the 
interface. 
They are represented by Dirac field operators $\psi^a_i({\vec x})$, with 
$SU(3)_c$ color index $a=1,2,3$, for red, green and blue, and flavour 
index $i=1,2,3$, for the three u,d,s light quarks. The three flavours are 
assumed, for simplicity, to be degenerate in mass. The matrices 
$\gamma^{\mu}$ are the usual 4D Clifford matrices and $C=i\gamma^0 
\gamma^2$ is the charge conjugation matrix.
The region outside of the ball is modeled, 
for simplicity, by a gas of free quarks. The interface
region between plasma and CS phases  is parameterized
by an effective order parameter that is 
proportional to the expectation value of the diquark condensate, 
$\Delta^{ab}_{ij} \propto \langle \psi^{a T}_i C\gamma^5 \psi^b_j 
\rangle$ in CS phase, and it is zero
in the plasma phase. We keep only a single tensor structure
\begin{equation}
\label{order}
\Delta^{ab}_{ij}({\vec x}) = \Delta({\vec x}) \left(\delta^a_i \delta^b_j 
- \delta^a_j \delta^b_i\right),
\end{equation}
relevant for CFL (Color Flavor Locking) phase. In this expression
$\Delta({\vec x})$ is treated as a   background field. In the superconducting 
phase $\Delta({\vec x}) = \Delta_{CS} \sim 100~$MeV is quite large and describes the  
energy gap in the spectrum of excitations. 
Outside the ball, where the system is treated as a plasma of free 
quarks, there is no Bose-condensation and $\Delta({\vec x}) = 
\Delta_{QP}=0$. 

In the limit of three massless flavours of quarks the symmetry of QCD 
interactions is enlarged to allow axial and vector  $SU(3)_A \times U(1)_A
\times SU(3)_V\times U(1)_B$ flavour rotations,
 in addition to color gauge transformations $SU(3)_c$. Notice 
that one of the diagonal matrices of $SU(3)_V$ is the generator of 
electromagnetic gauge transformations. The diquark condensate 
spontaneously breaks the symmetry group $SU(3)_c \times SU(3)_V\times SU(3)_A$ into the 
global $SU(3)$ diagonal subgroup of symmetry of lagrangian 
(\ref{lagrangian}), which locks color and flavour indices\cite{cs_r}. All the 
gauge bosons acquire masses $\sim |\Delta|$ via the Anderson-Higgs 
mechanism, except for a certain linear mix of the photon and one of the 
gluons that remains massless. Spontaneous breaking of the global symmetries 
  leads to formation of nine  pseudogoldstone bosons (  the octet of ``pions",   
 and   ``${\it \eta'}$" singlet, 
analogous to the   pseudoscalar mesons in the hadronic phase), 
and a single massless scalar corresponding to superfluid collective mode of the broken
 $U(1)_B$\cite{cs_r}.
These light, spin  zero fields can play an important role in transport properties,
however they do not carry baryon charge and, therefore, will be ignored in what follows.
\subsection{Quasiparticles in the superconducting phase.  }
The lagrangian density (\ref{lagrangian}) yields to the Dirac 
equation:
\begin{equation}
\label{eq1}
(i \partial_{\mu}\gamma^{\mu}- m + \mu\gamma^0)
\psi^a_i - \Delta^{ab}_{ij} C\gamma_5 {\bar \psi}^{b T}_j = 0.
\end{equation}
For $\Delta \neq 0$ the wavefunctions of quark fields and their
hermitian conjugates couple together and it is convenient to treat
the hermitian conjugate of equation (\ref{eq1}) as a second independent
equation:
\begin{equation}
{\bar \psi}^a_i (i 
\stackrel{\leftarrow}{\partial}_{\mu}\gamma^{\mu}+ m - 
\mu\gamma^0) - (\Delta^{ab}_{ij})^* \psi^{b T}_j C\gamma_5 = 0.
\end{equation}
The tensor  structure (\ref{order}) allows to decouple the set of Dirac 
equations in four sectors:
a) Three two-quarks channels (or 2SC sectors):
\begin{equation}
\label{2SC}
\left( 
\begin{array}{c}
u_{green} \\
d_{red} \\
0  \end{array} 
\right), \hspace{0.4in}
\left(
\begin{array}{c}
u_{blue} \\
0 \\
s_{red} \end{array}
\right), \hspace{0.4in}
\left(
\begin{array}{c}
0 \\
d_{blue} \\
s_{green} \end{array}
\right)
\end{equation}
b) One three-quarks channel (or CFL sector):
\begin{equation}
\label{CFL}
\left( 
\begin{array}{c}
u_{red} \\
d_{green} \\
s_{blue} \end{array}
\right).
\end{equation}
We discuss here the CFL sector, but the analysis of the three 2SC sectors 
is quite similar, \cite{Sadzikowski:2002ib}. 
In the CFL channel the set of Dirac equations can be written: 
\begin{equation}
\label{quarkIII}
\begin{array}{ccccccc}
(i \partial_{\mu}\gamma^{\mu} - m + \mu\gamma^0)
& u_{red} &  - & \Delta & C\gamma_5
& ({\bar d}_{green} + {\bar s}_{blue})^T& = 0, \\
(i \partial_{\mu}\gamma^{\mu} - m + \mu\gamma^0)
& d_{green}   &  - & \Delta & C\gamma_5
& ({\bar u}_{red} + {\bar s}_{blue})^T& = 0 \\
(i \partial_{\mu}\gamma^{\mu} - m + \mu\gamma^0)
& s_{blue}   &  - & \Delta & C\gamma_5
& ({\bar u}_{red} + {\bar d}_{green})^T& = 0.
\end{array}
\end{equation}
together with the hermitian conjugate expressions
\begin{equation}
\label{antiquarkIII}
\begin{array}{cccccccc}
&{\bar u}_{red} & (i
\stackrel{\leftarrow}{\partial}_{\mu}\gamma^{\mu} + m -
\mu\gamma^0) & - & \Delta^* & (d_{green} + s_{blue})^T & C\gamma_5 
& = 0, \\
&{\bar d}_{green} & (i
\stackrel{\leftarrow}{\partial}_{\mu}\gamma^{\mu} + m -
\mu\gamma^0) & - & \Delta^* & (u_{red} + s_{blue})^T & C\gamma_5 & 
= 0, \\
&{\bar s}_{blue} & (i
\stackrel{\leftarrow}{\partial}_{\mu}\gamma^{\mu} + m -
\mu\gamma^0) & - & \Delta^* & (u_{red} + d_{green})^T & C\gamma_5 & 
= 0.
\end{array}
\end{equation}
This set of six equations can be  decoupled: 
\begin{equation}
\label{quarkIIIb}
\begin{array}{cccccccc}
&(i \partial_{\mu}\gamma^{\mu} - m + \mu\gamma^0) 
\chi & - & 2 \Delta & C\gamma_5
& {\bar \chi}^T& = 0, \\
&{\bar \chi} (i
\stackrel{\leftarrow}{\partial}_{\mu}\gamma^{\mu} + m -
\mu\gamma^0 ) & - & 2 \Delta^* & \chi^T & C\gamma_5
& = 0,
\end{array}
\end{equation}
\begin{equation}
\label{quarkIIIc}
\begin{array}{cccccccc}
&(i \partial_{\mu}\gamma^{\mu} - m + \mu\gamma^0) 
\omega_1 & + & \Delta & C\gamma_5
& {\bar \omega}_1^T& = 0, \\
&{\bar \omega}_1 (i
\stackrel{\leftarrow}{\partial}_{\mu}\gamma^{\mu} + m -
\mu\gamma^0 ) & + & \Delta^* & \omega_1^T & C\gamma_5
& = 0,
\end{array}
\end{equation}
\begin{equation}
\label{quarkIIId}
\begin{array}{cccccccc}
&(i \partial_{\mu}\gamma^{\mu} - m + \mu\gamma^0) 
\omega_2 & + & \Delta & C\gamma_5
& {\bar \omega}_2^T& = 0, \\
&{\bar \omega}_2 (i
\stackrel{\leftarrow}{\partial}_{\mu}\gamma^{\mu} + m 
-\mu\gamma^0 ) & + & \Delta^* & \omega_2^T & C\gamma_5
& = 0.
\end{array}
\end{equation}
for the three independent combinations of fields defined as 
follows, $\chi=u_{red}+d_{green}+s_{blue}$, 
$\omega_1=u_{red}-d_{green}$ and $\omega_2=u_{red}-s_{blue}$.
The three sectors are formally identical except for one important
aspect that was already noticed in \cite{Sadzikowski:2002ib}: 
in the sector (\ref{quarkIIIb}) the energy gap, that is, the energy 
thershold at which single particles can be excited in the superconductor,
is $2|\Delta|$, twice the energy gap in the other two sectors 
(\ref{quarkIIIc}) and (\ref{quarkIIId}). We limit our 
analysis  to a single sector (\ref{quarkIIIc}), so we omit the 
subscript for $\omega$, and introduce the usual notation   
$\alpha^i=\gamma_0\gamma^i$ to simplify the equations,
\begin{equation}
\label{quarkIII2b}
\begin{array}{cccccccc}
&(i \partial_t + i \partial_{i} \alpha^{i} - m\gamma^0 + \mu)
\omega & + & \Delta & C\gamma_5
& {\omega^{\dagger}}^T& = 0, \\
&\omega^{\dagger} (
i \stackrel{\leftarrow}{\partial}_{t} +
i \stackrel{\leftarrow}{\partial}_{i} \alpha^i +
m \gamma^0 - \mu ) & + & \Delta^* & \omega^T & C\gamma_5
& = 0,
\end{array}
\end{equation}
In the Dirac equations   we are dealing with, the fermionic fields are 
quantum operators. However, in what follows, we neglect many body  effects 
and treat $\omega\rightarrow \varphi(t,{\vec x}) $
and $\omega^{\dagger}\rightarrow \zeta^{\dagger}(t,{\vec x})
$ as c-functions describing the single particle states\footnote{Similar 
procedure was adopted in \cite{Sadzikowski:2002ib}, after \cite{Blonder}, 
where the c-functions were defined in terms of the corresponding 
one-particle matrix elements. 
This procedure is well justified when the chemical potential $\mu$ 
is large because the many body effects are strongly suppressed in this case.}, 
\begin{equation}
\label{quarkIII3b}
 \left(
\begin{array}{c}
\varphi(t,{\vec x}) \\ 
\zeta^{\dagger}(t,{\vec x})
\end{array}
\right) = 
\left( 
\begin{array}{ccccc}
\sum_s  u_s({\vec q})\hspace{0.01in} 
\alpha_s({\vec q}) \\
\sum_s  v^{\dagger}_s(-{\vec q})  
\beta^*_s({\vec q})
\end{array} 
\right)exp(-iEt + i{\vec q}\cdot {\vec x}),
\end{equation}
where $u_s({\vec q})$ and $v_s(-{\vec q})$ are Dirac spinors which 
describe particles and antiparticle (or holes), respectively, and
the subindex $s = \uparrow,\downarrow$ denotes the two possible components 
of spin. They obey the equations
\begin{equation}
\label{quarkIII4b}
\begin{array}{cccccccc}
&({\vec \alpha}\cdot{\vec q} + m\gamma^0 - \mu) u_s({\vec q}) & = & 
+\kappa_{\vec q} \hspace{0.03in} 
u_s({\vec q}), \\
&v^{\dagger}_s(-{\vec q}) ({\vec \alpha}\cdot{\vec q} - m\gamma^0 + \mu) & 
= & 
-v^{\dagger}_s(-{\vec q}) \hspace{0.03in} 
\kappa_{\vec q},
\end{array}
\end{equation}
where $\kappa_{\vec q}=\sqrt{{\vec q}^2+m^2}-\mu$.
The coefficients $\alpha_s({\vec q})$,$\beta_s({\vec q})$ are c-numbers in 
this  approach. They obey the Bogolubov - de Gennes 
equations. For $ \alpha_{\uparrow}({\vec q})$ and  
$\beta^*_{\downarrow}({\vec q})$,
\begin{equation}
\label{coeffA}
\begin{array}{cccccccc}
&E \alpha_{\uparrow}({\vec q}) & = & 
+\kappa_{\vec q} \hspace{0.03in} \alpha_{\uparrow}({\vec q}) & 
+ \Delta \beta^*_{\downarrow}({\vec q}), \\
&E \beta^*_{\downarrow}({\vec q}) & = & 
-\kappa_{\vec q} \hspace{0.03in} \beta^*_{\downarrow}({\vec q}) & 
+ \Delta^* \alpha_{\uparrow}({\vec q}), \\
\end{array}
\end{equation}
and   similar  equations for  $\alpha_{\downarrow}({\vec q})$ and 
 $\beta^*_{\uparrow}({\vec q})$,
\begin{equation}
\label{coeffB}
\begin{array}{cccccccc}
&E \alpha_{\downarrow}({\vec q}) & = &
+\kappa_{\vec q} \hspace{0.03in} \alpha_{\downarrow}({\vec q}) &
- \Delta \beta^*_{\uparrow}({\vec q}), \\
&E \beta^*_{\uparrow}({\vec q}) & = &
-\kappa_{\vec q} \hspace{0.03in} \beta^*_{\uparrow}({\vec q}) &
- \Delta^* \alpha_{\downarrow}({\vec q}). \\
\end{array}
\end{equation}
There are two possible solutions of the uniform equations  
(\ref{quarkIII3b}) for the superconductor, characterized by $|\Delta| \sim 
100$~MeV, \cite{Sadzikowski:2002ib}:
\begin{eqnarray}
\label{general}
\left(
\begin{array}{c}
\varphi(t,{\vec x}) \\ \zeta^{\dagger}(t,{\vec x})
\end{array}
\right) = 
\left(
\begin{array}{ccccccc}
&e^{+i\frac{\delta}{2}} \sqrt{\frac{E+\xi}{2E}} 
(A u_{\uparrow}({\vec q_1}) - B u_{\downarrow}({\vec q_1})) \\
&e^{-i\frac{\delta}{2}} \sqrt{\frac{E-\xi}{2E}} 
(A {v^{\dagger}}^T_{\downarrow}(-{\vec q_1}) + B 
{v^{\dagger}}^T_{\uparrow}(-{\vec q_1})) \\
\end{array}
\right) e^{-iEt + i{\vec q}_1\cdot {\vec x}}  
\nonumber \\
+ \left(
\begin{array}{ccccccc}
&e^{+i\frac{\delta}{2}} \sqrt{\frac{E-\xi}{2E}}
(C u_{\uparrow}({\vec q_2}) + D u_{\downarrow}({\vec q_2})) \\
&e^{-i\frac{\delta}{2}} \sqrt{\frac{E+\xi}{2E}}
(C {v^{\dagger}}^T_{\downarrow}(-{\vec q_2}) - D
{v^{\dagger}}^T_{\uparrow}(-{\vec q_2})) \\
\end{array}
\right) e^{-iEt + i{\vec q}_2\cdot {\vec x}},
\end{eqnarray}
where $\xi=\sqrt{E^2-|\Delta|^2}$ 
and ${\vec q}_{1,2}=
\pm\sqrt{(\mu \pm \xi)^2 - m^2}$ and $\delta = arg(\Delta/|\Delta|)$.
They describe single particle excitations in the superconductor. 

The qualitative features of the scattering process at the interface 
between CS and plasma phases can be understood without explicit
solving the equations. First of all, $|\Delta|$ describes a gap in the 
spectrum of excitations in the CS phase. This is easy to see if we take $E 
< |\Delta|$. In this case 
${\vec q}_{1,2}=\pm\sqrt{\mu^2 - m^2 -|\xi|^2 \pm2i|\xi|\mu}$ have imaginary
parts, which implies an exponential suppression   in the wavefunctions.
This means that a single quark with low energy cannot propagate
in the superconductor because there is no any kinematically available
state for such an excitatation. At such low energies the quarks 
from outside  can only penetrate into the
superconductor if they organize a Cooper pair, which can be excited
without surpassing the gap energy, and leaving a hole that propagates 
backward in the metal. This phenomenom is known in condensed 
matter physics as the Andreev reflection at the interface between a metal 
and a superconductor \cite{Andreev}. Andreev reflection is an important 
phenomenom \cite{Blonder} when the potential barrier at the 
metal-superconductor interface is very low and the penetrability 
of the barrier is high. Otherwise, the normal reflection of 
low-energy incident electrons at the interface is the dominant process, 
as we will show in the coming subsection.
\subsection{Physics at the interface.}

We proceed in this section to write down the general equations which 
describe the interface of a plasma of quarks and the color 
superconducting phase. First, as there is no spin flip in the 
scattering process, we limit ourselves by considering the case
$B=D=0$ corresponding to a single spin sector.
Furthermore, for the sake of simplicity we assume that
the background field $\Delta({\vec x})$ depends only on the $z$ spatial 
coordinate. Then, the wavefunctions $\varphi(z)$ and $\zeta^{\dagger}(z)$ 
are invariant over the perpendicular spatial plane and the scattering 
problem can be reduced to a problem in one dimension. We can further
simplify the problem while still capturing its essential features by 
considering a step-function background $\Delta(z)$ to separate the 
interface between the phase of free quarks/antiquarks
and the superconducting phase:
$\Delta(z) = 0, \mbox {if} \hspace{0.03in} z \leq 0$ and 
$\Delta(z) = \Delta_0 \sim 100\mbox{~MeV}, \mbox {if} \hspace{0.03in} z > 
0$. Then the set of equations can be solved separately in each of the two 
phases and the corresponding wavefunctions must be matched continuously at 
the interface $z = 0$. It must be noticed that the equations we need to
solve are first order in the spatial derivative, and therefore we do 
not impose continuity conditions on the first derivatives 
\cite{Sadzikowski:2002ib}. This is a technical difference with respect to 
the treatment presented in the classical reference \cite{Blonder} of the 
interface between ordinary metal and superconductor, where electrons are 
described in the non-relativistic framework of Schrodinger's quantum 
mechanics. 

In Sector I, which describes the free quarks/ antiquarks at $z < 0$ where
$\Delta = 0$, the wave function (\ref{general}) is a linear superposition, 

\begin{eqnarray} 
\hspace{-0.8in}
\label{incident}
\left(
\begin{array}{c}
\varphi_+(z) \\
 \zeta^{\dagger}_-(z)
\end{array}
\right) =
A 
\left(
\begin{array}{c}
u(k_1) \\
0
\end{array} 
\right) e^{i k_1 z} +
A' 
\left(
\begin{array}{c}
u(-k_1) \\
0
\end{array} 
\right) e^{-i k_1 z} +
B' 
\left(
\begin{array}{c}
0 \\
v^{{\dagger}^{T}}(-k_2)
\end{array} 
\right) e^{i k_2 z},
\end{eqnarray} 
of the wave functions for an incident particle (first term, 
proportional to $A$), a reflected particle (second term, proportional to 
$A'$) and a reflected antiparticle (last term, proportional to $B'$). 
Obviously, an identical analysis can be done for the scattering of an
incident antiparticle at the interface. The 
ratio $|A'|^2/|A|^2$ gives the probability for normal reflection at the 
interface; while $|B'|^2/|A|^2$ will give the probability for Andreev 
reflection. In the expression above $k_1 = \sqrt{(\mu_1 + E)^2 - m^2}$ 
and $k_2 = \sqrt{(\mu_1 - E)^2 - m^2}$, where we remind $E$ is the energy 
above the Fermi surface of the incoming quark, $m$ is its mass and $\mu_1$ 
is the quark chemical potential in the plasma. We have added a subindex to 
the quark chemical potential $\mu_1$ in the plasma phase to point out that 
it can be different from the chemical potential $\mu_2$ in Sector II, 
which describes the superconducting phase. 

In Sector II) at $z >0$ the wave transmitted through the interface is a 
linear superposition 

\begin{eqnarray} 
\hspace{-0.8in}
\label{transmitted}
\left(
\begin{array}{c}
\varphi_+(z) \\
 \zeta^{\dagger}_-(z)
\end{array}
\right) =
a
\left(
\begin{array}{c}
e^{+i\delta/2} \sqrt{\frac{E+\xi}{2E}} u(q_1) \\
e^{-i\delta/2} \sqrt{\frac{E-\xi}{2E}} v^{\dagger^T}(-q_1) 
\end{array} 
\right) e^{i q_1 z} +
b
\left(
\begin{array}{c}
e^{+i\delta/2} \sqrt{\frac{E-\xi}{2E}} u(q_2) \\
e^{-i\delta/2} \sqrt{\frac{E+\xi}{2E}} v^{\dagger^T}(-q_2) 
\end{array} 
\right) e^{i q_2 z}, 
\end{eqnarray} 
of a transmitted quasiparticle (first term, proportional to $a$) and a
transmitted anti-quasiparticle (second term, proportional to $b$). In this 
expression $q_{1,2} = +\sqrt{(\mu_2 \pm \xi)^2 - m^2}$ with $\xi = 
\sqrt{E^2 - |\Delta|^2}$. In the superconducting phase the chemical 
potential is much larger than the current masses of the quarks, $m \ll 
\mu_2$, which will be neglected in what follows.
At energies $E > |\Delta|$ larger than the gap 
in the spectrum of quasiparticles in the superconductor the ratios 
$|a/A|^2$ and $|b/A|^2$ give, respectively, the probability for an 
incident quark with energy $E$ to be transmitted as a quasiparticle or 
anti-quasiparticle into the superconductor. As we have explained in 
previous sections, at energies $E < |\Delta|$ lower than the gap in the 
spectrum of quasiparticles in the superconductor particles/antiparticles 
from the plasma cannot propagate in the superconductor. 
This is manifest in the solution (\ref{transmitted}) because the 
transmitted wave is not a propagatting plane wave, but it decays 
exponentially.

The solutions in Sector I) and Sector II) must match continuously at
the interface $z = 0$, which demands

\begin{equation}
\label{normal}
A u(k_1) + A' u(-k_1) = a e^{i \delta/2} \sqrt{\frac{E + \xi}{2E}} u(q_1) 
+ b e^{i \delta/2} \sqrt{\frac{E - \xi}{2E}} u(q_2)
\end{equation} 

\begin{equation}
\label{Andreev}
B' v^{\dagger}(-k_2) = a e^{-i \delta/2} \sqrt{\frac{E - \xi}{2E}} 
v^{\dagger}(-q_1) 
+ b e^{-i \delta/2} \sqrt{\frac{E + \xi}{2E}} v^{\dagger}(-q_2)
\end{equation} 

The spinors $u(p)$ and $v^{\dagger}(p)$ are defined by equation 
(\ref{quarkIII4b}) and they can be explicitly solved in the form:

\begin{eqnarray}
u(p) = 
\left(
\begin{array}{c}
(m + \sqrt{p^2 + m^2})^{1/2}\ \chi \\
\frac{p \sigma_3}{(m + \sqrt{p^2 + m^2})^{1/2}}\ \chi
\end{array}
\right),
\end{eqnarray}

\begin{equation}
\hspace{-0.5in}
v^{\dagger}(-p) = 
\left(
{\bar \chi}^{\dagger} (m - \sqrt{p^2 + m^2} + 2\mu)^{1/2},\
-{\bar \chi}^{\dagger} \frac{p \sigma_3}{(m - \sqrt{p^2 + m^2} + 
2\mu)^{1/2}}
\right),
\end{equation}
in terms of 
\begin{eqnarray}
\chi = \left( 
\begin{array}{c} 
1 \\ 
0 
\end{array} 
\right)
\hspace{1.0in}
{\bar \chi}^{\dagger} = \left(1, 0 \right).
\end{eqnarray}

When these expressions are substituted in (\ref{normal}), (\ref{Andreev})
we obtain the final set of four equations that fix the four parameters
$A'$, $B'$, $a$ and $b$ in terms of the normalization constant $A$:

\begin{eqnarray}
\label{first}
\hspace{-0.8in}
e^{-i \delta/2} \left(A + A' \right) \left(m + \sqrt{k_1^2 + m^2} 
\right)^{1/2} = 
a \sqrt{\frac{E+\xi}{2E}} (q_1^2)^{1/4} + b \sqrt{\frac{E-\xi}{2E}} 
(q_2^2)^{1/4},
\end{eqnarray}

\begin{eqnarray}
\label{second}
\hspace{-0.8in}
e^{-i \delta/2} \left(A - A' \right) \frac{k_1}{\left(m + \sqrt{k_1^2 + 
m^2} \right)^{1/2}} = 
a \sqrt{\frac{E+\xi}{2E}} (q_1^2)^{1/4} + b \sqrt{\frac{E-\xi}{2E}} 
(q_2^2)^{1/4}, 
\end{eqnarray}

\begin{eqnarray}
\label{third}
\hspace{-0.8in}
\nonumber
e^{+i \delta/2} B' \left(m - \sqrt{k_2^2 + m^2} + 2\mu_1 
\right)^{1/2} = \\
\hspace{+0.6in}
a \sqrt{\frac{E-\xi}{2E}} \left(2\mu_2 - \sqrt{q_1^2}
\right)^{1/2} + b \sqrt{\frac{E+\xi}{2E}} \left(2\mu_2 - \sqrt{q_2^2}
\right)^{1/2},
\end{eqnarray}

\begin{eqnarray}
\label{fourth}
\hspace{-0.8in}
\nonumber
e^{+i \delta/2} B' \frac{k_2}{\left(m - \sqrt{k_2^2 + 
m^2} + 2\mu_1\right)^{1/2}} = \\
\hspace{+0.6in}
a \sqrt{\frac{E-\xi}{2E}} \frac{q_1}{\left(2\mu_2 - \sqrt{q_1^2}
\right)^{1/2}} + b \sqrt{\frac{E+\xi}{2E}} \frac{q_2}{\left(2\mu_2 - 
\sqrt{q_2^2} \right)^{1/2}},
\end{eqnarray}
where in the r.h.s of the equations, which describes Sector II, we have 
neglected the current masses of the quarks, as discussed above. These 
equations completely define the interations at the interface. 

\subsection{Small energies, $E < \Delta$.}

In this subsection we solve the set of equations  
(\ref{first})-(\ref{fourth}) in the setup that appropriately
describes the interface of a very diluted and cold plasma of quarks 
inciding with low energies $E \ll |\Delta| \simlt \mu_2$ on
the much dense superconducting medium of the balls. 
The analysis of this section is relevant for the understanding of the 
phenomenology of the balls and anti-balls in the cold universe, at 
temperatures much below $T_{QCD} \sim 160$~MeV. Although in this 
environment the CS balls and anti -balls coexist with a dilute gas of 
hadrons, rather than free quarks, the simpler description of the plasma 
phase as a diluted gas of quarks makes easier to understand the basic 
features of the interaction of real hadrons with the CS matter.

We specifically want to solve the equations (\ref{first})-(\ref{fourth})
when the incident quark from the metal is non-relativistic $k_{1,2} \simeq
\sqrt{2m\ E} \ll m$. This description corresponds to an incident massive
constituent quark with mass $m \simeq 300$~MeV. In this case 
the first two equations can be simplified in the form

\begin{equation}
\label{firstB}
\hspace{-0.7in}
e^{-i \delta/2} \left(A + A' \right) \sqrt{2m} + o(\frac{k_1}{m}) \simeq 
a \sqrt{\frac{+i|\Delta|}{2E}} (q^2_1)^{1/4} +
b \sqrt{\frac{-i|\Delta|}{2E}} (q^2_2)^{1/4} 
\end{equation}

\begin{equation}
\label{secondB}
\hspace{-0.7in}
e^{-i \delta/2} \left(A - A' \right) \frac{k_1}{\sqrt{2m}} + 
o(\left[\frac{k_1}{m}\right]^2) \simeq 
a \sqrt{\frac{+i|\Delta|}{2E}} (q^2_1)^{1/4} +
b \sqrt{\frac{-i|\Delta|}{2E}} (q^2_2)^{1/4} 
\end{equation}
which imply:

\begin{equation}
\left(A + A' \right) \sqrt{2m} \simeq \left(A - A' \right) 
\frac{k_1}{\sqrt{2m}} + o(\left[k_1/m\right]^2).
\end{equation}
and, therefore: 

\begin{equation}
A = - A' + o(\left[k_1/m\right]^2).
\end{equation}

The coefficient $|A'/A|^2 \simeq 1$ is precisely the probability for 
normal refelection of an incident low energy quark at the interface with 
the superconducting medium. Total probability conservation then implies 
that the probability for Andreev reflection in the analyzed setup is 
largely suppressed, $|B'/A|^2 \simeq o(\left[k_1/m\right]^2)$. As 
we have discussed in the previous section this result implies, due to 
time-reversal invariance, that the probability of low energy quarks 
annihilating at the interface of a ball of antiquarks in color 
superconducting phase is also largely supressed, 
$o(\left[k_1/m\right]^2)$.

The situation we just  described corresponds to the interface of color 
superconducting balls (or anti-balls) and a system of free
constituents quarks. As we mentioned, the relevant degrees of freedom 
outside the ball are not really the constituent quarks, but rather, 
hadrons made of confined quarks.
However, the arguments we presented above should convince the reader that
total reflection happens at the hadron-superconductor interface due to 
kinematic reasons not very sensitive to the confinement phenomena. 
Moreover, quantum numbers of individual quarks in a 
hadron do not exactly match those needed to form Cooper pairs in the 
superconductor, which implies that Andreev reflection of the hadron would 
demand a suppressed coherent collaboration of many quarks. This discussion 
explains why we claimed in the previous section that usual baryons 
colliding with the ball or antiball with a small  energy  $v/c\sim 
10^{-3}$,  will be mainly reflected by the ball/antiball. 

Many-body effects, which were completely ignored in these calculations,   
are not expected to change significatively the result of complete 
reflection of quarks and antiquarks off balls that we 
have described. Indeed, the annihilation of a quark with an antiquark coupled 
in a Cooper pair could be successful if the energy released from the 
annihilation is immediately transmitted to the second quark from the 
Cooper pair. In this case the second quark can receive sufficient energy
to overcome the gap and propagate as a quasiparticle in the superconductor.
The probability of this to happen is expected to be quite small because it 
involves interactions between many particles and, furthermore, the total
energy available from the current masses of the annihilating quark and 
antiquark in the dense CS phase is just above or maybe even below the gap 
threshold as we argued in Section II D. The precise estimate of this 
contribution requires the  use of quantum field theory
methods where the many body effects are properly taken into account.
Such estimates are beyond the scope of the present work,
and shall not be considered here. Indeed these details will only slightly change the
phenomenology of the balls/antiballs of CS matter/antimatter, but they will
not change the most profound implications of the cosmological scenario introduced
here. We also do not consider in this work, for the same reason, 
the physical interface region which  is probably  a mixture of nuclear and
color superconducting phases. The analysis of scattering in this case 
could be a very complicated problem. However, we expect that the 
noticeable suppression of the rate of matter 
antimatter anihilation still happens when the latter is stored in compact 
and dense objects of CS phase due to both a reduction in the cross section 
of the interaction and the reduction in the total flux of particle
antiparticle collision events. 

\subsection{Large energies, $E \simgt \Delta$}

In the text below we present some results on physics at the interface of 
superconducting and normal matter at large energies comparable to the 
gap. Such a study is not directly relevant for the the main purpose of 
this paper, which is the understanding of the phenomenology of the QCD 
balls (anti -balls) at the present epoch when the  Universe is cold and 
energies of particles surrounding the balls are very small. 
However, the study of these processes may give some insights about the 
mechanism of formation of the color superconducting regions during (or 
shortly after) the cosmological QCD phase transition (at $T \simeq 
\Delta$). Be adviced that we do not attempt to fully address the problem 
of formation of the balls, which would necessary include the analysis 
of the  non-perturbative dynamics of strong interactions during the QCD 
phase transition. Such analysis is beyond the scope of the present work. 

It seems natural to think that the ball
with the ground state to be the diquark condensate can form in a region 
where there already exists a large baryon density, that is, where $\mu$ 
is already large. 
Obviously, anti-balls are more prone to form in regions 
where there exists a large anti-baryon  density. The formation of large 
fluctuations of baryon number during the cosmological QCD phase 
transition has been discussed in a number of papers. For 
example, large fluctuations could occur if a first order phase transition 
takes place, see \cite{Sumiyoshi} where some applications 
to Big Bang Nucleosynthesis have  also been considered.  In those papers 
it was shown that the fluctuations in baryon number in the quark-gluon 
plasma during the QCD phase transition have a tendency to 
accumulate large number of baryons and become denser with time. Some 
authors even have suggested that the large fluctuations in baryon charge 
could result in the formation of the `strangelets' of quark matter originally
introduced in \cite{Witten}. 
Large baryon number fluctuations could also occur due to the axion related 
physics when the domain wall network with strong CP violation  
decays\cite{qcdball}, provided that the universe is already C asymmetric 
due to weak interactions. This effect could also lead to the separation of 
baryon/antibaryon charges. Finally, we have observed in sub-section II C 
that even in a closely baryon symmetric $B \simeq 0$, but largely C and CP 
asymmetric, universe at the onset of the QCD phase transition there can exist 
large chemical potentials $\mu_i$ of different quark subspecies in the 
asymptotically free plasma (only constrained to $\sum_i \mu_i \simeq 0$). 
In this context lumps of quark matter would naturally form at the end of the 
phase transition in quark sectors with large positive chemical potential 
$\mu_i > 0$, while antilumps of quark antimatter would form in the sectors of 
quarks with large negative chemical potential $\mu_j < 0$.

We do not have much to add to this difficult subject on the dynamics of 
the QCD phase transition. It is not our goal to describe the nature of the 
fluctuations of baryon number which eventually may seed the formation of 
balls. Our remark is the following: all Sakharov's conditions are satisfied 
when a fluctuations with a large baryon number in CS phase is formed, see
below, and, thus, the condensed region could continue to grow in size by 
accumulating the quarks coming from the plasma of free quarks. 
Indeed, one can suggest a simple model to account for this phenomenon by 
considering the ball surrounded by the dense plasma of quark matter with 
relatively large $\mu$. As we already mentioned, the set of equations that 
describe such a situation of the scattering of quarks at the interface has 
been solved previously in \cite{Sadzikowski:2002ib} in the context 
of the physics of neutron stars. There the transmission and reflection 
coefficients of quarks and holes (at high densities holes, rather than 
antiquarks, are the relevant quasiparticles) were calculated and it was 
shown that at energies $E \simgt |\Delta|$ there would be a net current 
of baryon charge through the interface,

\begin{equation}
\label{flux}
\hspace{0.9in}
 j_z = 2\mu_2
\frac{2\sqrt{E^2-|\Delta|^2}}{E+\sqrt{E^2-|\Delta|^2}}.
\end{equation}
Indeed, it was shown that in these conditions incident quarks from the surrounding
plasma can only be transmitted through the interface or Andreev reflected. 
Both processes contribute to the net transport (\ref{flux}) of baryon number 
into the condensed phase:
$\Delta B = +1$ when the incident quark is transmitted into the bulk of 
the supreconducting phase and $\Delta B = +2$  when the incident quark is 
reflected into the plasma as a hole with baryon charge B = -1. 
Therefore, Andreev reflection (largely suppressed at energies $E \ll |\Delta|$) 
could be an important mechanism that leads 
to the growth of balls/anti-balls, which eventually leads to matter 
antimatter separation, at temperatures (energies) $T \simlt |\Delta|$.

\section{Mechanism of Separation of Baryonic Charges}
The main point of this work is that formation of the cosmological dark matter 
and baryon asymmetry are closely related phenomena and originated from the 
same physics during the QCD phase transition. Therefore, the mechanism of 
formation of the dark matter is essentially the same physical process 
which produces the baryon-antibaryon separation. The mechanism how a chunk 
of dense matter (which is identified with the dark matter) is formed 
during the QCD phase transition might include new particles or fields or
might require a strong first order phase transition, but those are 
questions that shall not be addressed in detail here. 
We simply assume that such kind of objects made of condensed quark matter 
can be formed. Our goal here is to discuss some general requirements 
which should be satisfied to have a succesful separation mechanism of the 
baryon charges which leaves a net positive baryon number in the ordinary 
hadronic phase.

\subsection{Sakharov's Criteria}
All three Sakharov's criteria \cite{Sakharov} are satisfied (in a looser 
sense) during the formation of the balls of CS phase, without the need to 
introduce any new physics  beyond the standard model (except for, maybe, 
a solution for the strong CP problem in QCD). Indeed, 

1.The diquark condensate $\langle \psi^T C \gamma^5 \psi \rangle \neq 0$ 
forms in a local spatial region, or in a sector of quarks, where the chemical
potential $\mu_i$ is large. The net quark density breaks C and CP symmetries. 

2.The diquark condensate $\langle \psi^T C \gamma^5 \psi \rangle \neq 0$ 
formed in the CS phase spontaneously breaks  the baryon symmetry.

3.The dynamics of the QCD phase transition may provide the non-equilibrium
conditions when the balls of the condensed phase are formed, if the transition
does not proceed adiabatically;
futhermore, the diquark condensate $\langle \psi^T C \gamma^5 \psi 
\rangle \neq 0$ form within a local region or quark sector of large chemical
potential, which violates CPT symetry. 

It has been known for a while that these three ingredients (notice in 
particular spontaneous, rather than explicit, breaking 
of the baryon symmetry) can be responsible for a mechanism of charge 
separation (see e.g. \cite{Widrow}  and review paper\cite{Dolgov1}, 
where some simple toy models were discussed to explain the phenomenon of 
charge separation), which do not generate a net baryon number.

Given these three ingredients, lumps of condensed quarks, as well as, 
antilumps of condensed antiquarks can both form in cosmologically
large comoving volumes. In order to produce a separation of charges that 
hides a net excess of antibaryons over baryons in these balls and 
antiballs, and leaves an equal but positive excess of hadrons over 
antihadrons, one more ingredient is necessary:

4. Cosmological dynamics before or at the onset of the QCD phase transition 
must develop large C and CP asymmetries homogeneously over the whole comoving 
visible universe. 

The large cosmological C, CP asymmetries are stored in large chemical
potentials $\mu_i$ in the different quark sectors (constrained to $\sum_i \mu_i =
0$, if the net baryon number of the universe is zero $B = 0$). The development
of these net quark densities can proceed before the QCD phase transition due
to independent processes, see below. Obviuosly, in order to develop net quark
densities these C and CP violating processes must proceed in 
out-of-equilibrium conditions.

\subsection{Hierarchy of spatial scales}

We must make some specific emphasis on the spatial correlation scales of 
the sources of violation of the different symmetries due to 
the formation of the condensate. We will discuss four fundamentally 
different spatial scales: the QCD scale, $\sim T^{-1}_{QCD}$, determined 
by the fundamental physics involved in this scenario; the typical scale of 
the balls, $\sqrt[3]{B} T^{-1}_{QCD} \sim 10^{-3}$~cm, where $B \sim 
10^{18}$ up to $B \simlt 10^{40}$ for the configurations discussed in 
\cite{Madsen:2001fu, Lugones:2003un, qcdball}; the comoving 
Hubble horizon $H^{-1}_{QCD} \sim 30$~km at the QCD phase transition; and, 
finally, the present Hubble horizon $H^{-1} \ll H^{-1}_{QCD}$. 

In the picture we advocate in this work, the universe carries zero total 
baryon number $B = 0$ but it is not invariant under C or CP transformations, 
because it has an excess of baryons in the hadronic phase (visible matter) 
and an excess of antibaryon number stored in the CS phase (dark matter). 
Such cosmological C and CP asymmetries have a very large correlation 
length $\sim H^{-1}$ comparable to the size of the present horizon, which 
is the typical scale where it is observed a homogeneous excess of hadronic 
baryons $n_B = B({\widetilde n}_{\bar B} - {\widetilde n}_B) = 10^{-10} 
n_{\gamma}$. On the other hand, the size of the objects where baryon
symmetry is spontaneously broken by the Bose-condensate is of the order of the
size of the balls $\sim \sqrt[3]{B}~T_{QCD}^{-1}$. This scale, although 
macroscopically large in comparison with the QCD scale $\sim T_{QCD}^{-1}$, 
is still very small in comparison with the comoving Hubble horizon  
$H^{-1}_{QCD}$ and, therefore, much smaller than Hubble horizon at the 
present time $H^{-1}$. Therefore, if the universe was globally C and CP 
symmetric on the Hubble scale at the onset of the cosmological QCD phase 
transition, equal number of balls and antiballs would be formed in 
the comoving Hubble volume leaving no net baryon number in the hadronic phase. 
On the contrary, in presence of large and homogeneous C and CP 
asymmetries over cosmological scales (condition 4 in the previous subsection), 
the excess in the number density of 
anti-solitons over the number density of solitons is naturally of order 
one, ${\widetilde n}_{\bar B}-{\widetilde n}_B \sim {\widetilde n}_{\bar 
B}, {\widetilde n}_B$, as we assumed in our estimation of the ratio 
$\Omega_{DM}/\Omega_B$ in Section 2. This excess fixes in fact the number 
density of remnant baryons which are left in the hadronic phase, without 
any need of fine tuning.
 
What could be then the source of the cosmological C and CP asymmetries?
The important observation needed here to answer this question is that 
cosmological C and CP asymmetries must not necessarily be produced at the 
same instant of the QCD phase transition but could, instead, have been 
generated at some earlier stages by any other independent mechanism. In 
particular, the source of the cosmological C asymmetry is very natural 
because C is largely violated by weak interactions and, therefore, there 
is no reason to expect that the universe would be C invariant at the 
onset of the QCD phase transition. The origin of the large scale CP 
asymmetry of the universe, on the other hand, is still unknown. It is 
thought it must rely on physics beyond the standard model during the very 
early history of the universe, because the only known CP phase in the CKM 
matrix seems to be unable to do the work. We remark that
any new CP cource in extensions of the standard model (say, 
supersymmetric phases, neutrino interactions,...) could be fit to generate 
a large cosmological CP asymmetry before the QCD phase 
transition in the scenario that we are introducing. Other possibility to 
consider in this context is that cosmological CP asymmetry may be related 
to the same mechanism which eventually solves the so-called strong CP problem.
To this day the preferred solution to the strong CP problem is the dynamics of
the hypothetical axion field. 
At temperature $T \simeq  T_{QCD}$ the {\it strong} CP phase has
not yet relaxed to its ground state and thus might be of order unity, 
$\theta(T_{QCD}) \sim 1$, so that CP is largely violated.
If this mechanism is supposed to generate an homogeneous asymmetry over 
the whole visible universe it is specifically required that the initial 
value $\theta(T_{QCD})$ is the same everywhere in the entire observed 
Universe. This will occur, for example, if the Universe underwent 
inflation  during or after the Peccei-Quinn symmetry breaking at $T \sim 
10^{15}$~GeV, which is the standard assumption in the axion- related 
physics. This explicit source of CP violation relaxes to zero $\theta 
(T=0) =0$ soon  after the QCD phase transition is completed and the chiral condensate
is formed. At this point 
the axion field is settled at its minimum. 
In conclusion: a wide variaty of possible mechanism could drive the 
universe to become C and CP globally asymmetric before or during the QCD 
phase transition while the total baryon number remains equal to zero.
Our proposal of charge asymmetric cosmological separation of phases is not 
commited to any of these or other possible source of CP violation.

\subsection{ Estimation of the baryon excess $n_B/n_{\gamma}$ }

We want to introduce the discussion of this section with a paragraph
from the textbook {\it The Early Universe}, by E.~Kolb and M.~Turner 
\cite{Kolb} where the idea of baryon-antibaryon separation at the QCD 
scale (rather than generation of a net baryon number) is explicily 
mentioned, 

``{\it In a locally baryon symmetric universe nucleons and antinucleons 
remain in chemical equilibrium down to a temperature of $\sim 22 MeV$, 
when $n_B/n_{\gamma}=n_{\bar B}/n_{\gamma}\simeq 7 \times 10^{-20}$, a 
number that is 9 orders of magnitude smaller than the observed value of
$n_B/n_{\gamma}$. In order to avoid the annihilation catastrophe
an unknown physical mechanism would have to operate at a temperature
greater than 38 MeV, the temperature when
$n_B/n_{\gamma}=n_{\bar B}/n_{\gamma}\simeq 8 \times 10^{-11}$
and separate nucleons and antinucleons.}''\footnote{The value of the 
baryon to photon ratio stated in this paragraph is, in fact, 
an old estimation. The value reported by the WMAP collaboration 
$n_B/n_{\gamma}\sim 6 \times 10^{-10}$ is an order of magnitude larger. 
Therefore, the temperature at which the mechanism of baryon separation 
should operate need to be somewhat larger $\sim 41 MeV$ than  $38 MeV$ 
stated in this paragraph.}

If there is a mechanism of segregation of quarks and antiquarks (into hadronic  
and color superconducting phases ) during or immediately after the QCD 
phase transition, and the universe is largely C and CP asymmetric already 
at  that time,  it leaves an excess of antiballs over balls of order 
one  ${\widetilde n}_{\bar B}-{\widetilde n}_B \sim {\widetilde n}_{\bar 
B}, {\widetilde n}_B$. The same mechanism produces an excess of 
hadrons over anti-hadrons of order $n_B-n_{\bar{B}} = B ({\widetilde 
n}_{\bar B}-{\widetilde n}_B )$. This excess is preserved until today as a 
net remnant density of hadronic baryons once the hadron antihadron 
annihilation has been completed when the temperature reaches $22 MeV$, 
$n_B-n_{\bar{B}}\simeq n_B$. Therefore, the theoretical calculation of the 
present ratio $n_B/n_{\gamma}$ in the hypothetical scenario that we are 
discussing is reduced to the calculation of the corresponding 
time (temperature $T_{form}$) when the balls/antiballs complete their 
formation. This temperature would be determined by many factors: 
transmission/reflection coefficients, evolution of the balls,
expansion of the universe, cooling rates, evaporation rates, maybe 
dynamics of the axion domain wall network, etc. All these effects are, in 
general, of the same order of magnitude. Therefore, a precise theoretical 
calculation of $T_{form}$ in this context is a very difficult task. 

To reproduce the precise observational value of the baryon number 
in the hadronic phase 
\begin{equation}
\label{ratio}
\eta \equiv  \frac{n_B - n_{\bar B}}{n_{\gamma}} \simeq 
\frac{n_B}{n_{\gamma}} \sim 10^{-10},
\end{equation}
within this scenario, the magnitude $T_{form}$ must be precisely
\begin{equation}
\label{observations}
 T_{form} \simeq 41 ~MeV \hspace{0.6in} (observation),  
\end{equation} 
according to the relation $n_B \sim  
exp\left(-\frac{m_N}{T_{form}}\right)$ and the methods described long ago 
\cite{Kolb}.
It should be noticed that the ratio (\ref{ratio}) is very sensitive to the 
precise value of the temperature $T_{form}$: small variations in the 
value (\ref{observations}) by a few MeV would change the ratio 
(\ref{ratio}) by several orders of magnitude. This should not be 
interpreted as a need for fine tuning. Indeed, it is only a consequence 
of the fact that the cosmic time scale, measured by the temperature of 
the universe, is very long compared with the characteristic QCD time 
scale.

With this point in mind, it is clear that any theoretical estimation of
$T_{form}$ must be obtained with high precission if we want to confront it
with the precise observational data. Unfortunately, such theoretical 
calculation is not feasible at this time. A very rough estimation of 
$T_{form}$ is possible, it can be easily obtained by noticing that the 
balls/antiballs become completely opaque 
\footnote{It is a simple exercise to estimate the number of collisions 
between ordinary hadrons and balls/antiballs of CS phase in a Hubble time
as a function of the temperature $T \simlt T_{form}$, as we did in Section II D 
for the average number
densities in the Galaxy at present time, and show that annihilation events
are already suppressed immediately after the balls complete their formation.}
for the baryon charge in both directions for incident particles with energies 
much lower than the gap $\Delta$. Independently, we know that the BCS type 
phase transition from quark gluon phase to color superconductivity
takes place at temperature $T_c\simeq 0.6\Delta$ \cite{cs_n,cs_r}. 
For the standard value of the energy gap $\Delta =100 MeV$, $T_c \sim 
60 ~MeV$. The typical energies of particles at this temperature will be  
also of the same order. Therefore, one should expect that when temperature 
drops by some factor $\sim 1/2$ or so, the number of particles with 
relatively high energy capable to overcome the gap barrier will be tiny. 
Then, most of the particles will be reflected at such temperature. We 
expect that at this point the balls complete their formation period,
and  the thermodynamical equilibrium of the balls with the environment
will be settled. Therefore, the temperature of formation $T_{form}$ can be 
roughly estimated to lie in the interval,
\begin{equation}
\label{T}
30 ~MeV\simeq \frac{1}{2} T_c \simlt T_{form} \simlt T_c \simeq 60 ~MeV,
\end{equation} 
which should be compared with the ``observational'' value 
(\ref{observations}). The factor $1/2$ in this estimate is, of course, a 
quite arbitrary numerical factor which accounts for a suppression of the 
density of particles with sufficient  energy to surpass the energy gap 
$\Delta$. Even this rough and quite arbitrary estimation can only predict
a value for $\eta$ within a window of several orders of magnitude.

Notwithstanding the lack of precission of our estimation, there is an 
essential feature that makes this scenario physically attractive: the 
baryon to photon ratio is fixed in this scenario by the scale $T_{form}$
at which the mechanism of separation operates, while in most other 
suggested scenarios the models generally lack a natural scale in the 
problem that fixes the ratio $\eta$. 
 
\section{Discussion}

We have discussed a cosmological scenario where the universe is largely C 
and CP asymmetric, but carries zero baryon number. Large amounts of quarks
and antiquarks would be stored in very heavy chunks of matter or
antimatter in the color superconducting phase that would have
formed during the cosmological QCD phase transition.
We argue that the process of formation can leave an excess of baryons 
over antibaryons in the hadronic phase.

We have studied possible phenomenological constraints on this scenario
and have concluded that the scenario is  not ruled out and even not tightly 
constrained by available data. In particular, current constraints on
antimatter in the universe would not directly apply to our scenario 
because chunks of antimatter do not easily annihilate with the normal
(hadronic) matter. This is a consequence of the well-known features of the
interaction of normal matter at the interface with a superconductor and, 
esentially, the small volume occupied by the balls/antiballs and their extremely 
low number density. 

The chunks of dense matter would contribute to the 
``nonbaryonic" dark matter of the universe, in spite
of their QCD origin, because the baryon charge   stored in the diquark condensate
would not be available for nucleosynthesys. Therefore the baryon charge
locked in  the chanks of dense matter does  not contribute
to $\Omega_B h^2 \simeq 0.02$. 
Direct searches of of non-topological solitons carrying net baryon number 
has been reported in \cite{Arafune}. These observational data neither 
can rule out the range of parameters $B \simgt 10^{17}$ and $M \sim B m_N$, 
in which the balls of condensed quarks/antiquarks would be termodynamically 
stable \cite{Lugones:2003un,qcdball}.

The most profound consequences of the scenario are formulated in section
{\bf II B} and section {\bf IV C}: 

$\bullet$ We have shown in section {\bf II B} that the ratio 
$\Omega_{DM}/\Omega_B \simgt 1$ can be naturally understood as a direct 
  consequence of the underlying QCD physics,
and it is related to the fact that   both contributions
are originated at the same instant during the QCD phase transition. As 
it is known
 this  ratio  is very difficult to understand 
if both contributions to the energy density of the universe do not have 
the same origin.  

$\bullet$ In section {\bf IV C} we have shown that the fundamental ratio 
(\ref{ratio}) can be naturally understood without any fine tuning 
parameters as a direct consequence of the underlying QCD physics.
This ratio is determined by the temperature $T_{form}$ (\ref{observations})
when the balls 
complete their formation. This temperature falls exactly into the 
appropriate range (\ref{T}) of values where the baryon 
density can assume its observed value (\ref{ratio}).

Therefore, the ``exotic", dense color superconducting phase in QCD, might 
be a much more common state of matter in the Universe  than the ``normal" 
hadronic phase we know. In conclusion, qualitative as our arguments are, 
they suggest that baryogenesis can proceed at the QCD scale, and might be 
tightly connected with the origin of the dark matter in the Universe.
The scenario proposed in this paper offers a framework wherein to carry 
out calculations of cosmological parameters $\Omega_{DM}/\Omega_B$ as 
well as $\eta\equiv n_B/n_{\gamma}$. Both parameters offer well-defined 
testable predictions, although the present lack of theoretical 
and experimental knowledge on the QCD phase transition makes impossible 
for now to lay down this predictions more precisely than we have done in 
this paper.

In spite of the intrinsic difficulties to test directly the ideas 
introduced in this paper, it would be very intersting to check how
can they be probed by a laboratory type experiment in the spirit of 
the Program Cosmology in the Laboratory(COSLAB). Over the last few years 
several experiments have been done to test ideas drawn from cosmology and 
astrophysics(see \cite{volovik} and web page\cite{COSLAB} of the latest 
COSLAB meeting for further details). Ideally, the first test of the ideas 
introduced in this paper should be the confirmation of the
features of the interaction of hadrons at the interface with a color 
superconducting phase discussed in section 3. We suggest that these 
features might be tested in the scattering of low energy positrons and 
electrons by a  conventional superconductor in some regime which would be 
analogous to the cosmological environment.

This scenario with no doubt leads to important consequences for cosmology 
and astrophysics, which have not been explored yet. In particular,
the recent detection\cite{Teplitz} of the seismic event with epilinear
(in contrast with a typical epicentral ) sources may  be related to a 
soliton-like {\it dark} object, which could be the very dense balls. 
Also, the cuspy halo problem in dwarf galaxies might be related to some 
kind of {\it interacting} \cite{Cen} or {\it annihilating} cold dark matter 
in the denser regions of galaxies and clusters,
which could be  related to the balls discussed in this work.
Last, unexpected excess of photons measured by a number of collaborations 
may also be related to the balls. In particular, $511$~KeV line 
emission from the galactic bulge \cite{astro1,astro2} could be related to 
the excess of positrons (electrons) surrounding the QCD anti-balls 
(balls). Also, the excess of $100$~MeV photons could be related
to the decay of goldstone modes propagating in the bulk of the balls.
We expect many other consequences which have not been explored yet.

\section*{Acknowledgments}
We are  thankful to  Sasha Dolgov for useful discussions, critical 
comments and  nice remarks. We are also thankful to other 
participants of   the ``Baryogenesis Workshop", (MCTP, Ann-Arbor, June 
2003)  where this work was   presented. One of us (AZ) thanks Michael Dine, 
Lev Kofman, Slava Mukhanov and Alex Vilenkin for discussions and comments. We 
are also thankful to Michael Peskin for discussions during his visit to 
Vancouver. DO also wants to thank T.~Pereg-Barnea, J.C.~Cuevas and J.~Oaknin 
for explaining to him important concepts.

This work was supported in part by the National Science and Engineering
Research Council of Canada. The financial support of
  Michigan Center for Theoretical Physics is greately appreciated.

\end{document}